%% file: scifimapaper.tex
\documentclass[twocolumn,showpacs,preprintnumbers,amsmath,amssymb,prd,superscriptaddress]{revtex4}

\usepackage{graphicx}
\usepackage{dcolumn}
\usepackage{bm}

\begin{document}


\title{Measurement of the quasi-elastic axial vector mass in neutrino-oxygen interactions}

\input{author}

\date{\today}

\begin{abstract}
\input{abstract}
\end{abstract}

\maketitle

\input{introduction}

\input{crosssections}

\input{experiment}

\input{analysis}

\input{results}

\input{conclusion}

\bigskip

\begin{acknowledgments}
\input{acknowlegements}
\end{acknowledgments}



%
%

\bibliography{scifimapaper}

\end{document}

%% file: author.tex
\newcommand{\BCN}{\affiliation{Institut de Fisica d'Altes Energies, Universitat Autonoma de Barcelona, E-08193 Bellaterra (Barcelona), Spain}}
\newcommand{\BU}{\affiliation{Department of Physics, Boston University, Boston, Massachusetts 02215, USA}}
\newcommand{\UBC}{\affiliation{Department of Physics \& Astronomy, University of British Columbia, Vancouver, British Columbia V6T 1Z1, Canada}}
\newcommand{\UCI}{\affiliation{Department of Physics and Astronomy, University of California, Irvine, Irvine, California 92697-4575, USA}}
\newcommand{\SACLAY}{\affiliation{DAPNIA, CEA Saclay, 91191 Gif-sur-Yvette Cedex, France}}
\newcommand{\CNU}{\affiliation{Department of Physics, Chonnam National University, Kwangju 500-757, Korea}}
\newcommand{\DU}{\affiliation{Department of Physics, Dongshin University, Naju 520-714, Korea}}
\newcommand{\DUKE}{\affiliation{Department of Physics, Duke University, Durham, North Carolina 27708, USA}}
\newcommand{\GENEVA}{\affiliation{DPNC, Section de Physique, University of Geneva, CH1211, Geneva 4, Switzerland}}
\newcommand{\UH}{\affiliation{Department of Physics and Astronomy, University of Hawaii, Honolulu, Hawaii 96822, USA}}
\newcommand{\KEK}{\affiliation{High Energy Accelerator Research Organization(KEK), Tsukuba, Ibaraki 305-0801, Japan}}
\newcommand{\HIR}{\affiliation{Graduate School of Advanced Sciences of Matter, Hiroshima University, Higashi-Hiroshima, Hiroshima 739-8530, Japan}}
\newcommand{\INR}{\affiliation{Institute for Nuclear Research, Moscow 117312, Russia}}
\newcommand{\KOBE}{\affiliation{Kobe University, Kobe, Hyogo 657-8501, Japan}}
\newcommand{\KOR}{\affiliation{Department of Physics, Korea University, Seoul 136-701, Korea}}
\newcommand{\KYO}{\affiliation{Department of Physics, Kyoto University, Kyoto 606-8502, Japan}}
\newcommand{\LSU}{\affiliation{Department of Physics and Astronomy, Louisiana State University, Baton Rouge, Louisiana 70803-4001, USA}}
\newcommand{\MIT}{\affiliation{Department of Physics, Massachusetts Institute of Technology, Cambridge, Massachusetts 02139, USA}}
\newcommand{\MIYAGI}{\affiliation{Department of Physics, Miyagi University of Education, Sendai 980-0845, Japan}}
\newcommand{\NIIGATA}{\affiliation{Department of Physics, Niigata University, Niigata, Niigata 950-2181, Japan}}
\newcommand{\OKAYAMA}{\affiliation{Department of Physics, Okayama University, Okayama, Okayama 700-8530, Japan}}
\newcommand{\OSAKA}{\affiliation{Department of Physics, Osaka University, Toyonaka, Osaka 560-0043, Japan}}
\newcommand{\ROME}{\affiliation{University of Rome La Sapienza and INFN, I-000185 Rome, Italy}}
\newcommand{\SNU}{\affiliation{Department of Physics, Seoul National University, Seoul 151-747, Korea}}
\newcommand{\SOLTAN}{\affiliation{A.~Soltan Institute for Nuclear Studies, 00-681 Warsaw, Poland}}
\newcommand{\TOHOKU}{\affiliation{Research Center for Neutrino Science, Tohoku University, Sendai, Miyagi 980-8578, Japan}}
\newcommand{\SB}{\affiliation{Department of Physics and Astronomy, State University of New York, Stony Brook, New York 11794-3800, USA}}
\newcommand{\TUS}{\affiliation{Department of Physics, Tokyo University of Science, Noda, Chiba 278-0022, Japan}}
\newcommand{\KAM}{\affiliation{Kamioka Observatory, Institute for Cosmic Ray Research, University of Tokyo, Kamioka, Gifu 506-1205, Japan}}
\newcommand{\RCCN}{\affiliation{Research Center for Cosmic Neutrinos, Institute for Cosmic Ray Research, University of Tokyo, Kashiwa, Chiba 277-8582, Japan}}
\newcommand{\TRIUMF}{\affiliation{TRIUMF, Vancouver, British Columbia V6T 2A3, Canada}}
\newcommand{\VAL}{\affiliation{Instituto de F\'{i}sica Corpuscular, E-46071 Valencia, Spain}}
\newcommand{\UW}{\affiliation{Department of Physics, University of Washington, Seattle, Washington 98195-1560, USA}}
\newcommand{\WARSAW}{\affiliation{Institute of Experimental Physics, Warsaw University, 00-681 Warsaw, Poland}}

\BCN
\BU
\UBC
\UCI
\SACLAY
\CNU
\DU
\DUKE
\GENEVA
\UH
\KEK
\HIR
\INR
\KOBE
\KOR
\KYO
\LSU
\MIT
\MIYAGI
\NIIGATA
\OKAYAMA
\OSAKA
\ROME
\SNU
\SOLTAN
\TOHOKU
\SB
\TUS
\KAM
\RCCN
\TRIUMF
\VAL
\UW
\WARSAW

\author{R.~Gran}\altaffiliation[Now at ]{University of Minnesota, Duluth}\UW
\author{E.~J.~Jeon}\SNU
\author{E.~Aliu}\BCN                
\author{S.~Andringa}\BCN 
\author{S.~Aoki}\KOBE 
\author{J.~Argyriades}\SACLAY 
\author{K.~Asakura}\KOBE 
\author{R.~Ashie}\KAM 
\author{F.~Berghaus}\UBC
\author{H.~Berns}\UW 
% ...
\collaboration{The K2K Collaboration}\noaffiliation

%% file: abstract.tex
The weak nucleon axial-vector form factor for quasi-elastic interactions
is determined using neutrino interaction data 
from the K2K Scintillating Fiber detector 
in the neutrino beam at KEK. 
More than 12,000 events are analyzed, of which 
half are charged-current quasi-elastic interactions 
$\nu_\mu n \rightarrow \mu^- p$
occurring primarily in oxygen nuclei.
We use a relativistic Fermi gas model for oxygen and 
assume the form factor 
is approximately a dipole with one parameter,
the axial vector mass $M_A$, and fit to the shape of the distribution
of the square of the momentum transfer from the nucleon to the nucleus.
Our best fit result for 
$M_A$ = 1.20 $\pm$ 0.12 GeV. 
Furthermore, this analysis includes updated vector form factors from recent 
electron scattering experiments and a discussion of the effects 
of the nucleon momentum on the shape of the fitted distributions.

%% file: introduction.tex
\section{introduction}

The structure of the nucleon, as measured both by electrons and neutrinos,
has been a subject of experimental study for decades.  The discovery of
neutrino oscillation and the availability of high precision electron scattering
measurements have renewed interest in the study of neutrino interactions 
on nuclei. 
Neutrinos offer unique information about the structure of the nucleon and the nucleus.
There are many experimental neutrino programs now running, under construction,
or being planned for the near future, all of which use nuclear targets
such as oxygen, carbon, aluminum, argon, or iron. 
Likewise, there has been significant progress
in the calculation of cross sections, nuclear corrections, and backgrounds
to specific processes.
Improvement of these models, supported by neutrino data, will be important
for the upcoming precision neutrino oscillation studies.

In this study we analyze distributions of the square of the 
four-momentum transfer
$Q^2 = -q^2= -(p_\mu - p_\nu)^2$ 
reconstructed from neutrino-oxygen interactions, where p$_\mu$ and p$_\nu$ are the 
momenta for the outgoing muon and incident neutrino.
Using data from the Scintillating 
Fiber (SciFi) detector in the KEK accelerator to Kamioka (K2K) neutrino beam,  
we fit for the value of the axial vector mass $M_A$, 
the single parameter in the axial 
vector form factor (assuming a dipole form) for quasi-elastic (QE) interactions. 
For QE interactions, this parameter is obtained only from neutrino-nucleus scattering 
experiments.  
This is the first such measurement
for oxygen nuclei, and we include a discussion of the effects of the 
oxygen nucleus and nucleon momentum distribution on the shape of the $Q^2$ 
distribution.

In the next section we briefly discuss the quasi-elastic form factors as well as
list the cross-sections for non-quasi-elastic processes and nuclear effects.
Then a section describes the K2K experiment, the neutrino beam, and the SciFi detector,
including the detector performance.  Following that are sections on the analysis technique,
and the results, which include detailed discussion of the major systematic effects.

%% file: crosssections.tex
\section{Cross section and form factor expressions}

\subsection{Quasi-elastic cross section}

The differential cross section $d\sigma/dq^2$ for neutrino quasi-elastic 
scattering ($\nu_\mu\;n\rightarrow\mu^-\;p$) is described in terms
of the vector, axial-vector, and pseudo-scalar form factors.  
The differential cross section\cite{LlewellynSmith:1972} is written as:
\begin{eqnarray}
\label{Eq:crosssection}
 \frac{d\sigma^{\nu}}{dq^2} & = &
 \frac{M^2G^2_F\cos^2\theta_c}{8\pi E^2_{\nu}} \times \\
\nonumber & & 
  \left[A(q^2)-B(q^2)\frac{s-u}{M^2}+C(q^2)\frac{(s-u)^2}{M^4}\right]
\end{eqnarray}
where, s and u are Mandelstam variables, (s-u) = $4ME_\nu + q^2 - m^2$,
m is the outgoing lepton mass, $M$ is the target nucleon mass, 
and $E_{\nu}$ is the neutrino energy.
$A(q^2)$, $B(q^2)$, and~$C(q^2)$ are:
\begin{eqnarray}
\nonumber   
A(q^2) & = & \frac{m^2-q^2}{4M^2} \left[  (4-\frac{q^2}{M^2})|F_A|^2 \right. \\
\nonumber
       &   & - (4+\frac{q^2}{M^2})|F^1_V|^2 - \frac{q^2}{M^2}|\xi F^2_V|^2(1+\frac{q^2}{4M^2}) \\ 
\nonumber
       &   &  \left. -\frac{4q^2F^1_V\xi F^2_V}{M^2} - \frac{m^2}{M^2}((F^1_V+\xi 
F^2_V)^2+|F_A|^2) \right], \\
\nonumber 
B(q^2) & = & \frac{q^2}{M^2}((F^1_V+\xi F^2_V)F_A), 
\nonumber \\
C(q^2) & = & \frac{1}{4}
        \left(|F_A|^2+|F^1_V|^2-\frac{q^2}{4M^2}|\xi F^2_V|^2\right).
\end{eqnarray}
In these expressions, 
the pseudo-scalar form factor $F_P$ is negligible for muon neutrino
scattering away from the muon production threshold and is not included.
$F_A$ is the axial vector form factor we will extract from the data.
$F^1_V(q^2)$ and $F^2_V(q^2)$ are the Dirac electromagnetic isovector form 
factor and the Pauli electromagnetic isovector form factor, respectively. 
These formulas also assume
the conserved vector current (CVC) hypothesis, which allows us to write $F^1_V$
and $F^2_V$ in terms of the well measured Sachs form factors
$G^P_E$, $G^N_E$, $G^P_M$, and $G^N_M$:
\begin{eqnarray}
\nonumber
F^1_V(q^2) & = & (1 - \frac{q^2}{4M^2} )^{-1}\Big[(G_E^P(q^2)-G_E^N(q^2)) \\ 
\nonumber & & -\; \frac{q^2}{4M^2} (G_M^P(q^2) - G_M^N(q^2))\Big], \\
\nonumber \xi F^2_V(q^2) & = & (1 - \frac{q^2}{4M^2})^{-1}\left[(G_M^P(q^2) - G_M^N(q^2))\right. \\ 
  & &  - \left.(G_E^P(q^2)-G_E^N(q^2))\right].
\end{eqnarray}
  
In this paper we use the updated
measurements of the Sachs form factors from \cite{Bosted:1995,BBA:2002}.
These new form factors have a significant effect on the extraction of 
$F_A$, compared to the previous dipole approximations.  For the range
of $Q^2$ of interest in this experiment, the updated values differ 
from the old form factors by up to $\pm$ 10\%.  
We present results with both the new and the old form factors in this paper.

We approximate the axial vector form factor $F_A$ as a dipole
\begin{equation}
 F_A(q^2) = -\frac{1.2720}{\left(1-(q^2/M^2_A)\right)^2},
\end{equation}   
which has a single free parameter, the axial vector mass $M_A$.
Previous studies show that this approximation is 
reasonable \cite{BNL:1990,Fermilab:1983,ANL:1982}.
The constant $F_A$($q^2$=$0$) = $g_A/g_V$ = 1.2720$\pm$0.0018 
is determined from neutron decay measurements\cite{RPPstat:2004}.
Because the Sachs form factors and other constants 
are precisely measured, the single parameter $M_A$ can be 
determined from quasi-elastic neutrino interaction data.

\subsection{Other cross sections}

For this analysis, approximately half of the data comes from non
quasi-elastic interactions, especially single pion events from the 
production and decay of the N$^\ast$ and $\Delta$ baryon resonances
within the nucleus.  
This background is described
by the NEUT neutrino interaction Monte Carlo simulation \cite{Hayato:2002} used by 
the K2K and Super-Kamiokande experiments. 
The resonance single pion events are from the model of Rein 
and Sehgal\cite{ReinSehgal:1981}.  

Additional backgrounds, less important for our beam energy
around 1.2 GeV, 
are also included.
Deep inelastic scattering is from GRV94\cite{GRV94} for the 
nuclear structure functions
with a correction described by
Bodek and Yang\cite{BodekYangNuInt01:2002}.
The software PYTHIA/JetSet\cite{PYTHIA} is used to generate these events.
This analysis takes the charged current coherent pion cross section to be zero 
following \cite{K2KCohPi:2005}, and include neutral current coherent pion interactions
as in Rein and Sehgal\cite{ReinSehgal:1983}.

\subsection{Nuclear Effects}

Equation \ref{Eq:crosssection} 
is the differential cross section for the {\em free} nucleon,
and must be modified to account for the effects of a nucleon bound in a
nucleus.  In the SciFi detector, the fiducial mass fractions are
0.700 H$_2$O, 0.218 Al, 0.082 HC,
with an error of $\pm$0.004.
Our neutrino interaction Monte Carlo treats the entire fiducial mass as
if it was made of H$_2$O; 
for targets other than a proton in hydrogen,
we use a uniform Fermi gas model with $k_f$ = 225 MeV/c for the nucleon
momentum and an effective binding energy of -27 MeV, which is appropriate
for oxygen.
The primary effect of this nucleon momentum distribution on 
the quasi-elastic events is a significant suppression at low $Q^2$
due to Pauli blocking and a smaller overall suppression of $\sim$2\% for 
the entire $Q^2$ distribution.  
The Fermi gas model is also applied to the non quasi-elastic interactions.

In addition to cross section effects, there are final-state interactions. 
The nucleus will cause reinteraction
or absorption of secondary pions and recoil protons and neutrons
in the neutrino interaction final state.  Our model for these 
reinteractions is described in \cite{Hayato:2002} with references.  
This will affect the observed distribution of the number of tracks.  
The resulting $\mu^-$
is also affected by the Coulomb interaction as it leaves the nucleus,
losing approximately 3 MeV, though this effect is implicitly included
in the Fermi gas parameters.
The above nuclear effects are discussed quantitatively in the results section.

%% file: experiment.tex
\section{Experiment}

\subsection{The beam and detectors}
The KEK to Kamioka (K2K)\cite{K2K:2001,K2K:2003,K2K:2005}
experiment is a long baseline neutrino oscillation measurement
in which a beam of neutrinos is sent from the KEK accelerator in 
Tsukuba, Japan to the underground Super-Kamiokande 
detector\cite{SuperK:NIM}.
The neutrinos pass through a set of near neutrino detectors 300 meters from the target,
after which they travel 250 km to Super-Kamiokande.
The analysis in this paper considers only neutrino interactions detected
in the Scintillating Fiber (SciFi) detector, one of the near detectors.

The wide-band neutrino beam at KEK is produced when 12 GeV protons hit an
aluminum target.  Two magnetic horns focus positively charged pions and kaons into
a 200 meter long decay pipe, where they decay to $\mu^+$ and $\nu_\mu$.  
The $\mu^+$ are absorbed by the beam dump plus approximately 100 
meters of earth between the decay pipe and near detector hall. 
The resulting neutrino energy is between 0.3 and 5 GeV 
and peaks at 1.2 GeV.
The contamination in this beam includes 1.3\% $\nu_e$ and
0.5\% anti-$\nu_\mu$, estimated from a Monte Carlo simulation of the 
beam.

\begin{figure}[htbp!]
\begin{center}
\includegraphics[width=8.5cm]{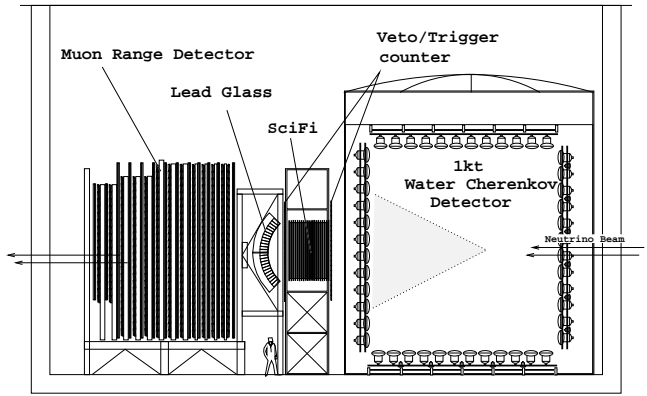}

\caption{
The arrangement of the near neutrino detectors at KEK.  The beam
comes in from the right and continues to Super-Kamiokande, 250 km away 
to the left.  The SciFi detector in on a stand in the middle.
}
\label{Fig.detectors}
\end{center}
\end{figure}

The near detector hall of the K2K experiment contains several detectors,
shown in Fig.~\ref{Fig.detectors}.
The first detector in the beam is the one-kiloton water Cerenkov detector.
This study uses data from the SciFi detector, which
is described in detail below.
Following SciFi is the location of the lead glass detector which was used
to measure the $\nu_e$ contamination in the beam.
The lead glass detector was removed in 2002 and in its place was a 
prototype for  a plastic scintillator (SciBar) detector.  Then in 
2003, the full SciBar detector~\cite{SciBar:2004} was installed, 
though data from this last
running period is not used in the present analysis.
Finally, there is a muon range detector (MRD)~\cite{MRD:2002}
which is used to estimate the momentum of the muons which escape the SciFi detector
from charged current neutrino interactions.
The MRD is also used to monitor the stability of the neutrino beam.

The prediction for the shape of the neutrino energy spectrum
of the K2K beam
has significant uncertainty, up to 20\% at higher energies.
This prediction uses a Sanford-Wang parameterization of hadron
production data and is verified using pion monitors downstream
from the target\cite{K2K:2001}.
This energy spectrum is measured
using data from the near detectors and is used as input to the oscillation analysis
\cite{K2K:2005}.  
The energy spectrum
analysis is a simultaneous fit to the muon momentum and muon angle distributions
from charged current interactions in 
the one-kiloton water Cerenkov detector, the SciFi detector, and the SciBar 
detector.  The free parameters in this fit are a scale factor for the flux 
in eight energy regions, a scale factor for non quasi-elastic events,
and many systematic error parameters specific to each detector.
For this paper, we will refer to the above procedure as the neutrino energy
spectrum measurement,  
and it defines the 
baseline Monte Carlo prediction for the SciFi detector data, prior to
any fitting for the axial-vector mass, and is used throughout the discussion 
and plots in this Sec. III. 
This default MC simulation also uses zero charged current coherent pion and
$M_A^{QE}$ = 1.1 GeV.
The resonance single-pion cross section also involves its own axial-vector term
with its own $M_A^{1\pi}$ = 1.1 GeV.  
The analysis described in Sec. IV and V is mostly independent from the energy spectrum
analysis, but uses a similar strategy.

\subsection{The SciFi detector}

\begin{figure}[htbp!]
\begin{center}
\includegraphics[width=8.0cm]{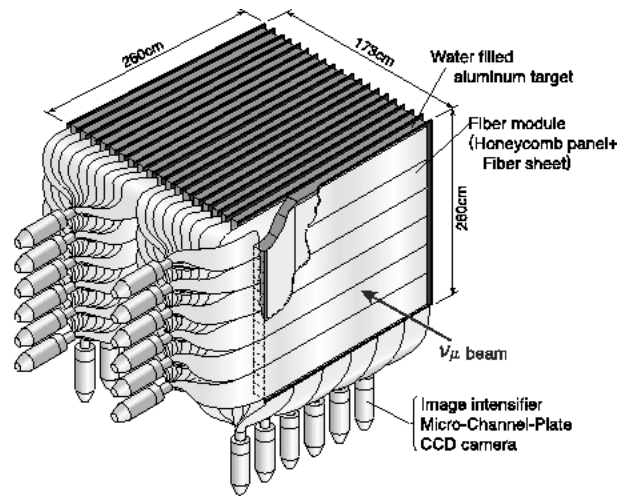}

\caption{
A schematic diagram of the SciFi detector.
}
\label{Fig.scifidetectors}
\end{center}
\end{figure}

The SciFi detector \cite{SciFi:2000,SciFi:2003}
consists of scintillating fiber
tracking layers between aluminum tanks filled with water. 
A schematic diagram is included in Fig.~\ref{Fig.scifidetectors}. 
There are a total of twenty 240 cm x 240 cm wide tracking layers, each of which
consists of fibers oriented to give the particle location in the horizontal 
and vertical direction.  These fibers are glued, one layer on each 
side, to a honeycomb panel which is 260 cm square.  The distance between 
two tracking layers is 9 cm.  Between the first and the twentieth layer
are nineteen layers of aluminum tanks whose walls are 0.18 cm thick with an 
interior thickness 6 cm filled with water.

The scintillating fibers have a diameter of 0.7 mm and are read out by 
coupling them to image intensifier tubes and CCD cameras.  
The image intensifier preserves
the position information of the original photo-electron.  At the final
stage, the light is recorded by a CCD camera.  A total of 24 of these
are used to read out 274,080 scintillating fibers.
To reconstruct which fibers
were hit, a one-to-one correspondence between the fibers and the position
of pixels on the CCD camera is obtained from periodic calibration using an
electro-luminescent plate.

To select charged-current events for this analysis, we require at least
one track start in the SciFi fiducial volume and extend to, and stop within, 
the MRD.
The fiducial volume has a mass of 5590 kg, 
includes the first through 17th tanks of water, 
and the reconstructed vertex must be within 110 cm from the center
of SciFi in horizontal and vertical directions.  

This requirement means
that all events selected for this analysis have hits in at least 
three tracking layers of SciFi.
There are also upstream
and downstream scintillator hodoscopes which are read out by photo-multiplier
tubes; we require a matching hit downstream and no hit upstream.
This allows us to veto the small number of muons which come from in-time 
muon generation in the rock and upstream material in the detector hall,
and also muons surviving from the target.  Cosmic ray muons are suppressed
by the beam timing requirement and are also negligible.

Tracks are reconstructed in the horizontal and vertical projections 
separately and then matched.  
The efficiency for reconstructing
muon tracks with hits in three SciFi tracking layers is $\sim$70\%, 
and rises to nearly 100\% for tracks that penetrate five or more layers.
Second tracks are required to produce hits in at least 
three SciFi layers, but there is no restriction on the maximum length.

Prior to 2002, all muons from SciFi are required to pass through 
and produce hits in segments of the 
lead glass detector, and these segments must match the location
of the reconstructed track seen in SciFi and the MRD.  
On average, they deposit around 0.4 GeV of energy in the lead glass, 
though only path length, and not pulse size, is used to estimate
the energy loss in this case.
In the K2K-II run period, muons traveling through the SciBar prototype lose
around 0.023 GeV of energy, though we do not require the track to 
pass through this detector.
Muons traveling through many layers in SciFi deposit up to 0.3 GeV of energy.

The Muon Range Detector is made of alternating layers of drift tubes 
and iron plates; 
the first detection layer is upstream of the first piece of iron.  
The first four layers have a thickness equivalent to 
about 0.14 GeV of energy loss each, and the remaining layers are 
twice as thick.  The muon momentum can then be estimated by 
calculating the muon's range from the interaction vertex.

\subsection{Data samples}

The data for this analysis are obtained from two running periods between 
November 1999 and June 2003.  The primary distinction between them is the
configuration of the Super-Kamiokande detector, though there were simultaneous
changes in the near detector configuration important for this analysis.
We refer to the first as the ``K2K-I'' period;
muons from neutrino interactions in SciFi pass through the lead glass 
detector on their way to the MRD.  For these data, we accept muons which
penetrate as little as one MRD detection layer, which corresponds to a 
muon momentum threshold of 675 MeV/c.  
The second running period is called ``K2K-IIa'' and has the prototype for the 
plastic scintillator detector SciBar \cite{SciBar:2004} 
in place of the lead glass.  
For K2K-IIa, we require that the muons produce hits
in the first {\em two} 
layers of the MRD, which gives a threshold of 550 MeV/c, in order to reduce
the contamination from pions reaching the MRD.  Data from the continuation of the 
K2K-II period are not used in this analysis.  In all cases, we require the muon
not exit the MRD, which results in a maximum muon momentum of 3.5 GeV/c.

When two tracks reach the MRD, 
the longest, most penetrating track is assumed to be the 
muon.  We have estimated using the MC that approximately 
2\% of these longest tracks are not the muon track,
and another 0.5\% of events were from neutral current interactions
which had no muon at all.

This analysis uses only one-track and two-track events.
Since quasi-elastic interactions will not produce such events,
the 3\% of events with three or more reconstructed tracks are discarded.
For one-track events, the recoil proton or a pion is absent or below threshold.  
The requirement of three layers for the second track corresponds to a 
threshold of 600 MeV/c proton momentum and 200 MeV/c pion momentum.

The MC simulation includes the rescattering
of protons, neutrons, pions, and other hadrons from the neutrino interaction final state
as they leave the nucleus.
The models for these final state interactions give good agreement with the 
number of tracks seen in SciFi, shown in Tab.\ref{Tab.numtracks}.
Our estimate of the uncertainty in the number of two-track events
due to the efficiency for finding the second track is 5\%, which translates
into $\pm$ 100 events for the K2K-I two-track sample.  
These events primarily migrate to or from the one-track sample.
The extreme case of zero nuclear final state interactions in the 
neutrino interaction MC leads to 20\% more events in the two-track 
category~\cite{Walter:NuInt01}.
Further discussion of these effects can be found in Sec. V of this paper. 

\begin{table}[htbp!]
\begin{tabular}{l|ccc}
sample & 1-track & 2-track & 3-track
\\ \hline K2K-I Data & 5933 & 2181 & 187
\\ K2K-I MC & 5920 & 2176 & 203
\\ \hline K2K-IIa Data & 3651 & 1344 & 148
\\ K2K-IIa MC & 3583 & 1396 & 136
\end{tabular}
\caption{Comparison between data and MC simulation of the number of reconstructed tracks 
observed in the SciFi detector for the K2K-I and K2K-IIa data samples.  
The MC is normalized to have the same
number of events as the data.  The estimated error in the number of two-track events due to
tracking efficiency in reconstructing short, second tracks is 5\%.}
\label{Tab.numtracks}
\end{table}

\begin{figure}[htbp!]
\begin{center}
\includegraphics[width=8.5cm]{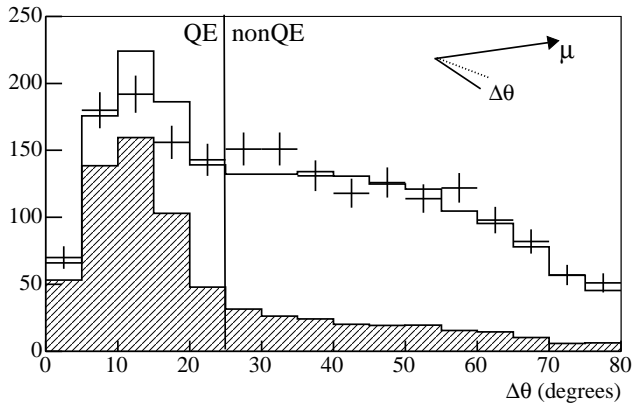}

\caption{
The distribution of $\Delta\theta$, the difference between the predicted second track angle and the observed angle for K2K-I data.  The histogram shows the Monte Carlo prediction, while the hatched region shows the QE fraction.  In this analysis, $\Delta\theta = 25$ degrees is used to separate a QE enhanced sample. The inset diagram shows
the definition of $\Delta\theta$.  
}
\label{Fig.DeltaTheta}
\end{center}
\end{figure}

For two-track events, we separate quasi-elastic
from non quasi-elastic events.  
Quasi-elastic interactions are a two-particle scattering process; 
the measurement of the muon momentum
and angle is sufficient information to predict the angle of the recoil
proton.  If the measured second track agrees with this prediction within
25$^\circ$, it is likely a QE event.  If it disagrees, then it becomes a 
part of the non-QE sample.  This is shown in Fig.~\ref{Fig.DeltaTheta},
where the inset diagram demonstrates the kinematic quantity 
$\Delta\theta$ = angle between predicted and measured 
second track angle with respect to the beam.
The quantity $\Delta\theta$ is plotted in this figure with the data and
the baseline MC normalized to the data.

The value for this $\Delta\theta$ cut is chosen to give
good separation between the QE and nonQE enhanced samples.
We used the Monte Carlo simulation to estimate 
the efficiency for detecting the QE events,
after all the cuts described above.  Also we have
estimated the purity of each sub-sample.  These are 
shown in Tab.~\ref{Tab.Efficiency}.  After these cuts,
the total number of events in each sample is given in Tab.~\ref{Tab.Events}.

\begin{table}[htb]
\begin{center}
\begin{tabular}{l|ccc|c}
& 1-track & \multicolumn{2}{c}{2-track} & Total \\
& & QE & nonQE & \\
\hline K2K-I &\hspace{0.5em}35 (63)\hspace{0.5em}&\hspace{0.5em}5 (63)\hspace{0.5em}&\hspace{0.5em}2 (17)\hspace{0.5em}&\hspace{0.5em}42\hspace{0.5em} \\
K2K-IIa\hspace{1em}&\hspace{0.5em}38 (61)\hspace{0.5em}&\hspace{0.5em}5 (61)\hspace{0.5em}&\hspace{0.5em}2 (15)\hspace{0.5em}&\hspace{0.5em}45\hspace{0.5em}
\end{tabular}
\end{center}
\caption{Total reconstruction efficiency [\%] for quasi-elastic interactions
in each data set, the portion of efficiency from each sub-sample, 
and the QE purity of each sample (in parenthesis, [\%]), estimated with the MC simulation.}
\label{Tab.Efficiency}
\end{table}

\begin{table}[htb]
\begin{center}
\begin{tabular}{c|cc|cc}
& \multicolumn{2}{c}{K2K-I} & \multicolumn{2}{c}{K2K-IIa} \\
  &  \hspace{0.05em}$Q^2 > 0.0$ \hspace{0.05em} 
  & $Q^2 > 0.2$ \hspace{0.1em} 
  & \hspace{0.05em} $Q^2 > 0.0$ \hspace{0.05em} 
  & $Q^2 > 0.2$ \hspace{0.1em} \\
\hline 1 track       & 5933 & 2864 & 3623 & 1659 \\
2 track QE    &  740 &  657 &  451 &  388 \\
2 track nonQE & 1441 &  789 &  893 &  478 \\
\hline Total                  & 8114 & 4310 & 4967 & 2525
\end{tabular}
\end{center}
\caption{Number of events in three event samples and two data periods 
for the SciFi detector.  Only events with reconstructed $Q^2 > 0.2$
(GeV/c)$^2$ are used for this $M_A$ measurement, and are shown in 
separate columns and described in Sec.~IV.}
 \label{Tab.Events}
\end{table}

\subsection{Muon momentum and angle distributions}

\begin{figure}[htbp!]
\begin{center}
\includegraphics[width=8.5cm]{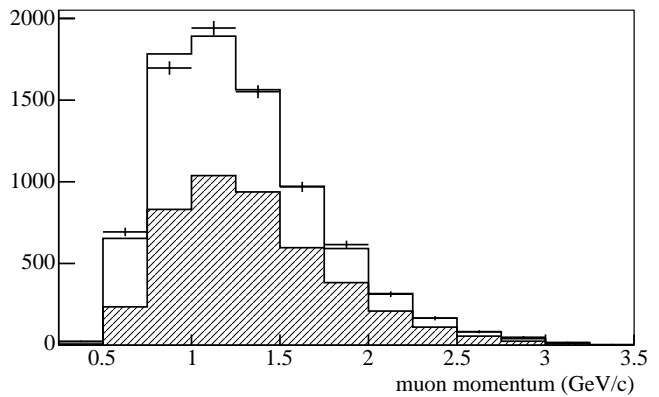}

\caption{Muon momentum distribution for all K2K-1 one-track and two-track events.
The QE fraction, estimated from the MC simulation, is shown as the shaded region.
The errors on the data are statistical only.
}
\label{Fig.pmu}
\end{center}
\end{figure}

\begin{figure}[htbp!]
\begin{center}
\includegraphics[width=8.5cm]{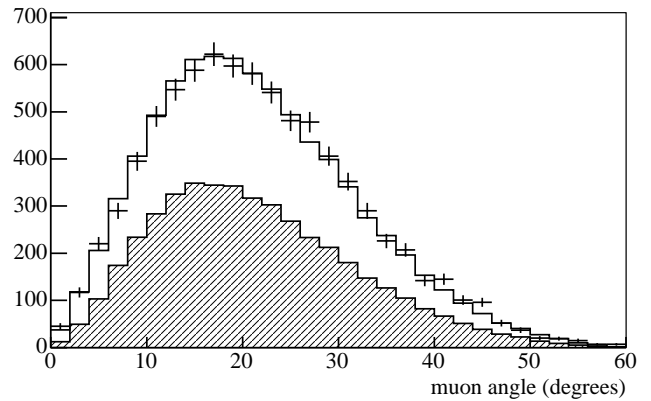}

\caption{Muon angle distribution for all K2K-I one-track and two-track events.
The QE fraction, estimated from the MC simulation, is shown as the shaded region.
Only statistical errors are shown.
}
\label{Fig.thetamu}
\end{center}
\end{figure}

An example of the muon momentum and muon angle distributions for the K2K-I data
along with the Monte Carlo prediction are shown in Fig.~\ref{Fig.pmu} and Fig.~\ref{Fig.thetamu}.
The MC distribution is normalized to the same number of events.

\begin{figure*}[htbp!]
\begin{center}
\includegraphics[width=5.8cm]{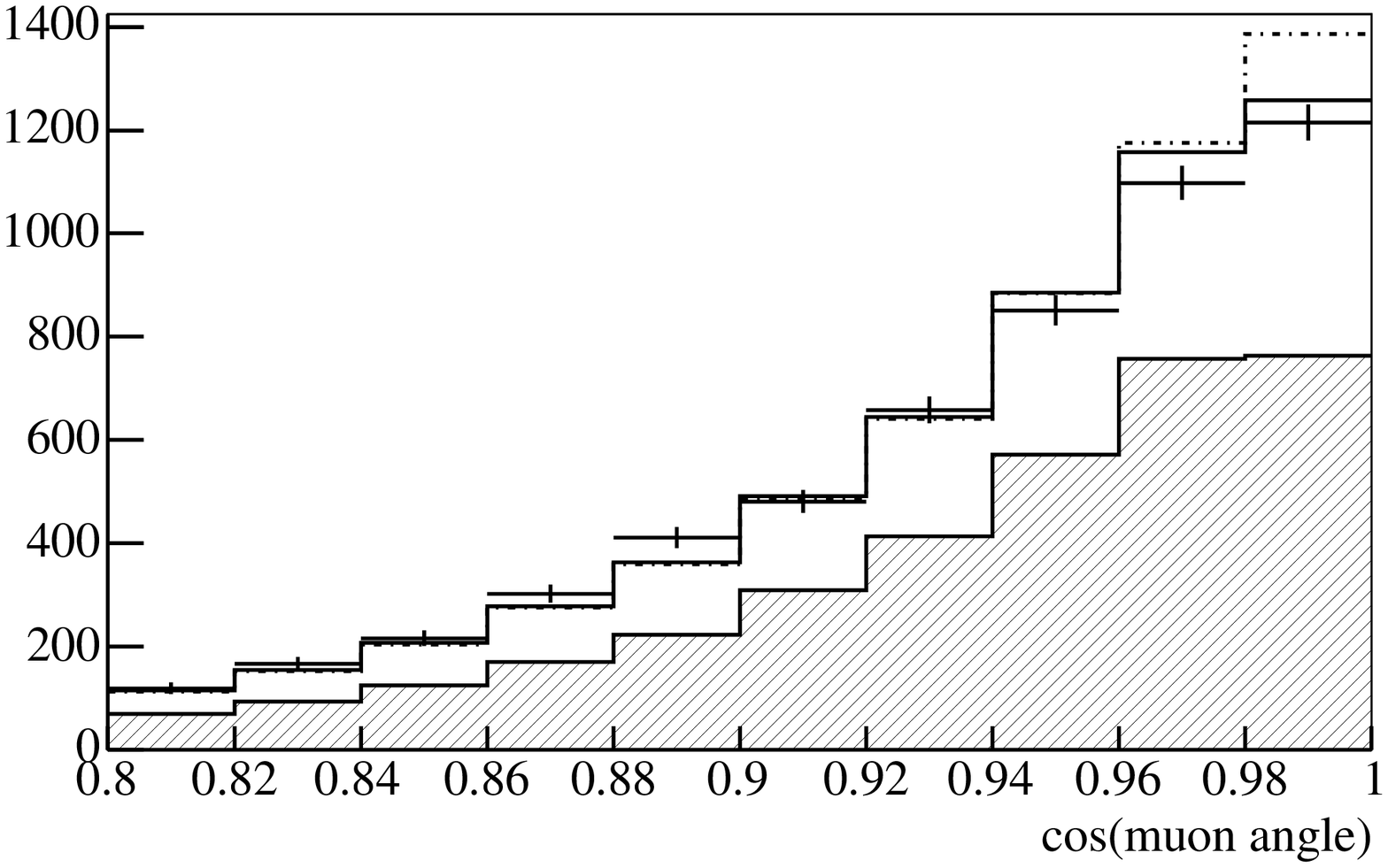}
\includegraphics[width=5.8cm]{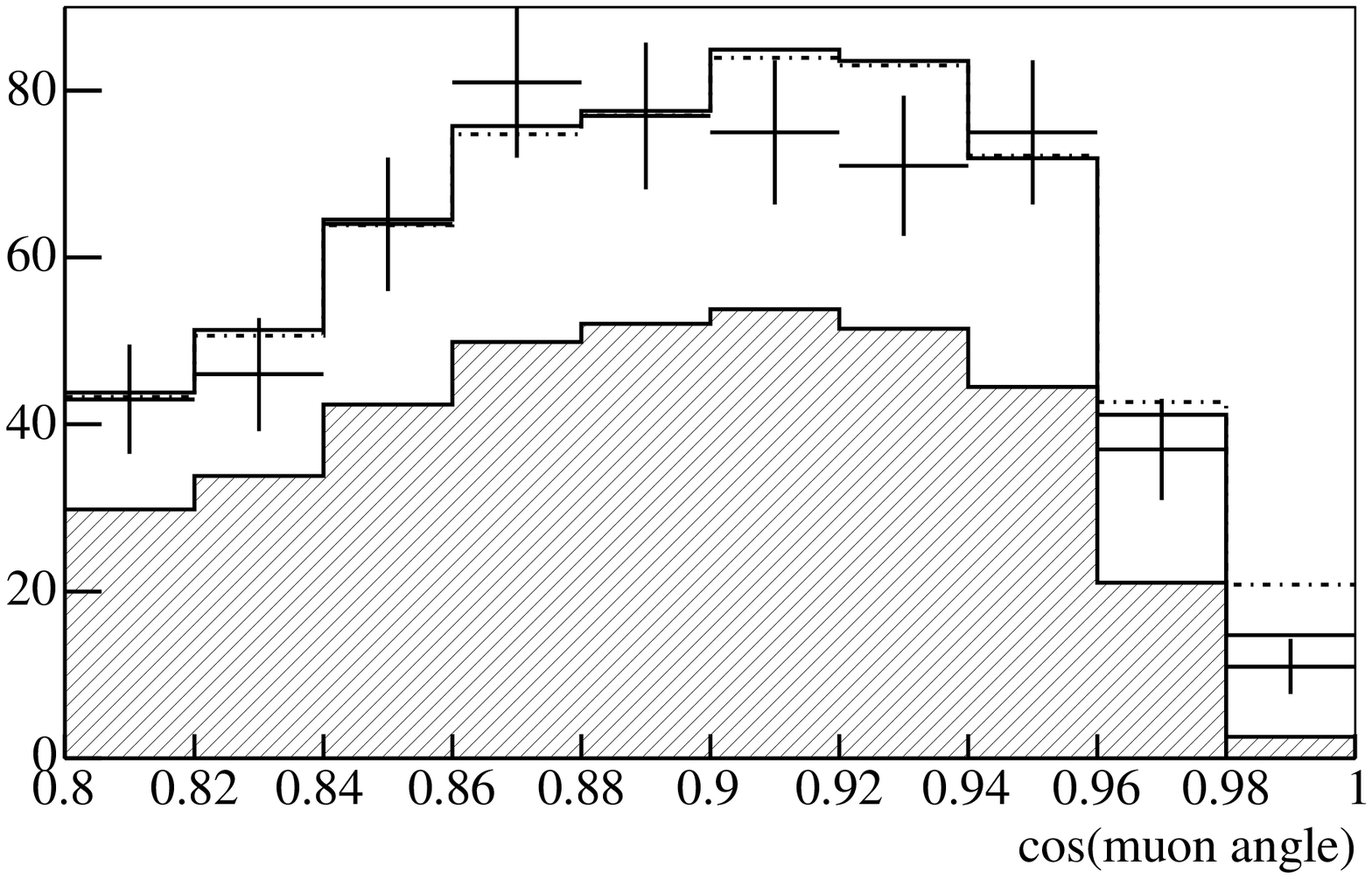}
\includegraphics[width=5.8cm]{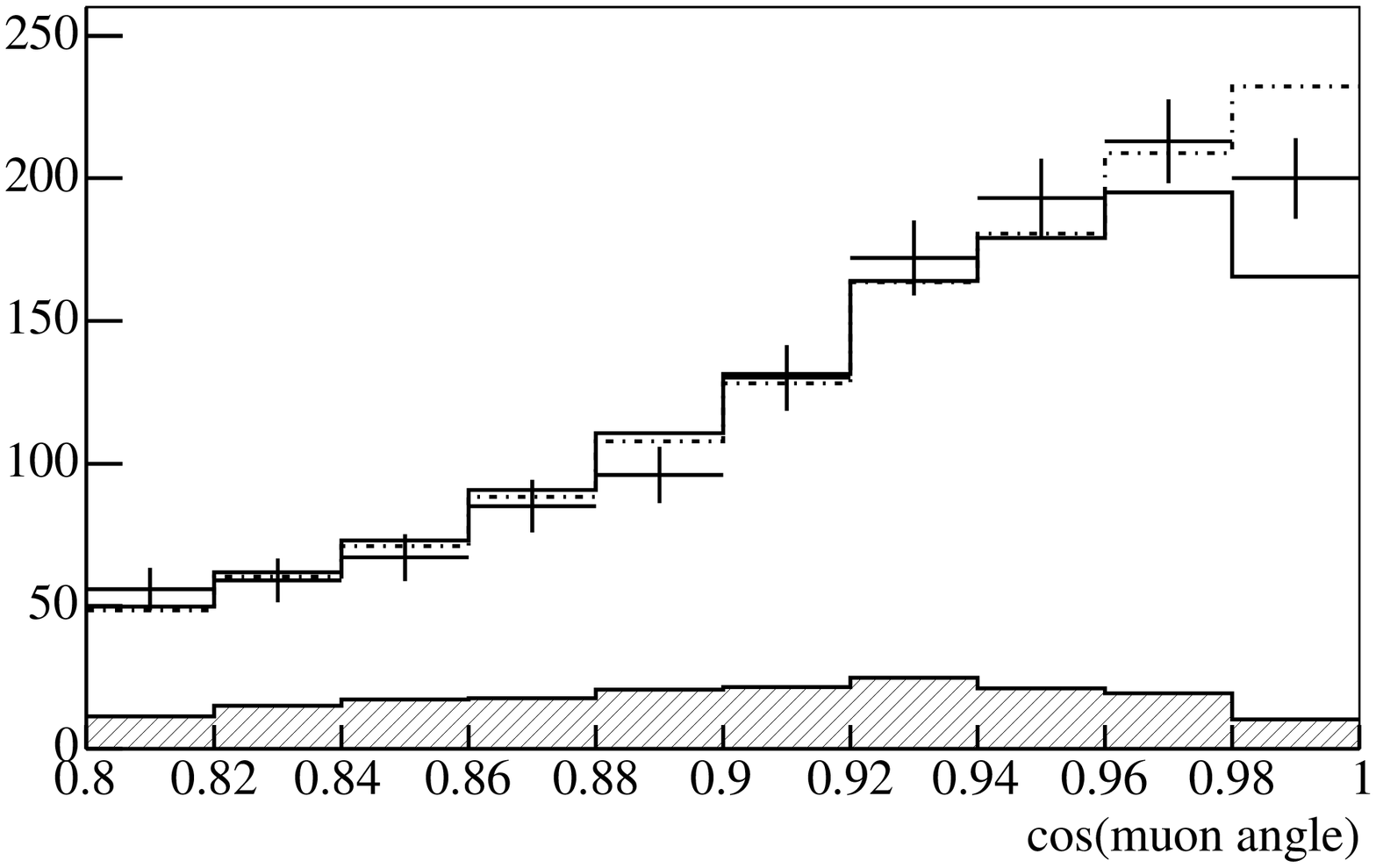}

\caption{Distribution of $\cos(\theta_\mu)$ for data and MC, showing the
data at small angle, which is also the low $Q^2$.  Left to right are the
one-track, two-track QE enhanced, and two-track nonQE enhanced samples. 
The top line is with
charged-current coherent pion, the second line without.  The shaded region
is the QE fraction, estimated from the MC simulation.  The distributions
are normalized to the total number of events when all three samples are combined.
}
\label{Fig.lowQ2}
\end{center}
\end{figure*}

We observed a deficit of events whose muon is
at angles near the direction of the beam compared to our 
Monte Carlo simulation;
this is also discussed in \cite{K2K:2005}.  
The discrepancy was observed
in all K2K near detectors, including SciFi, and is presumed to be from
some aspect of the neutrino interaction model.  The analysis of data 
from the SciBar detector \cite{K2KCohPi:2005} indicated that most, if not all of
this deficit is because there is too much CC coherent pion production in 
the MC.  
The SciBar data are consistent with zero charged-current coherent pion.
Examples of the disagreement from SciFi data are 
shown in Fig.~\ref{Fig.lowQ2}, with and without charged current coherent pion events.
The MC distribution with zero CC coherent pion is normalized to the 
same number of events.  
In this analysis, we assume there is zero CC coherent $\pi$ production.

%% file: analysis.tex
\section{Analysis}

\subsection{Calculating $Q^2$ and $E_\nu$}

The kinematics of the muon candidate, the longest track in our events,
are sufficient to estimate the energy of the neutrino $E_{\nu}^{rec}$
and the square of the momentum transfer $Q^2_{rec}$,
if the interaction is quasi-elastic.

\begin{eqnarray}
 E_{\nu}^{rec} & = & \frac{(m_N + \epsilon_B)E_\mu - 
                  (2m_N \epsilon_B + \epsilon_B^2 + m^2_\mu)/2}
                  {m_N + \epsilon_B - E_\mu + p_\mu \cos \theta_\mu} , \\
 \nonumber & & \\
 Q^2_{rec} & = & -q^2 = -2E_\nu(E_\mu - p_\mu \cos \theta_\mu) + m_\mu^2.
\label{Eq.Q2}
\end{eqnarray}

Here, $E_\mu$ and $p_\mu$ are the energy and momentum of the muon 
determined from the range, $\theta_\mu$ is the angle relative to the
beam direction, determined from the hits in the SciFi detector.  
Note that $E_\nu$ appears in the expression for $Q^2_{rec}$,
and here we use $E_\nu^{rec}$.  
The quantity $\epsilon_B$ = -27 MeV for oxygen is the
effective binding energy parameter from the Fermi gas model. 
The masses $m_N$ and $m_\mu$ are for the nucleon and the muon.
The resolution for $E_\mu$ is 0.12 GeV, due mainly to the 
MRD segmentation.
The resolution for $\theta_\mu$ is about 1 degree, but there is a
tail to this distribution.
The resulting value for $E_\nu$ resolution (for QE events) is 0.16 GeV and the
resolution for $Q^2$ is 0.05 (GeV/c)$^2$ also with a tail coming
from the measured angle. 
Finally, this formula assumes that the target neutron inside the nucleus 
is at rest, ignoring the nucleon momentum distribution for the event
reconstruction.  Fluctuations due to Fermi motion are about half the size of those
due to detector and reconstruction effects, and contribute only a small
amount to the reconstructed energy resolution.

It is important to note that these formulas are used for all events 
even though half the interactions are not quasi-elastic, because we
do not identify the interaction mode on an event-by-event basis, nor
is our beam at a fixed energy.
The reconstructed $E_\nu$ and $Q^2$ are systematically off for these
non quasi-elastic events:
$E_\nu^{rec}$ is low by $\sim$0.4 GeV 
and $Q^2_{rec}$ is low by $\sim$0.05 (GeV/c)$^2$.
However, all events are treated the same way, both data and
Monte Carlo events. Thus, the comparison of data and MC in the fit is valid,
but the distributions of the reconstructed values are affected by the 
non quasi-elastic fraction.

\subsection{Fit procedure}

After calculating $E_\nu^{rec}$ and $Q^2_{rec}$ for each event, 
the data are binned
in five $E_\nu^{rec}$ bins:  0.5 to 1.0, 1.0 to 1.5, 1.5 to 2.0, 2.0 to 2.5, 
and greater than 2.5 GeV.  The data are divided into $Q^2$ bins each of 
width 0.1 (GeV/c)$^2$.  To ensure there are at least five events in 
each bin, the smaller number of events at higher $Q^2$ are combined into 
a single bin.

The expectation for the number of the events in each bin is computed
from the Monte Carlo simulation
for different values of the axial-vector mass and some systematic error parameters.
%
We perform a maximum likelihood fit to the data by minimizing the 
negative of the logarithm of the likelihood which is based on Poisson
statistics for each bin.  In our case we use the modified form given
in the Review of Particle Physics \cite{RPPstat:2004}
\begin{eqnarray}
-2 \ln \lambda(\theta) = 2 \sum^N_{i=1} [ \nu_i(\theta) - n_i
+ n_i \ln (n_i / \nu_i(\theta))]
\end{eqnarray}
in which $\nu_i(\theta)$ and $n_i$ are the predicted and observed values
in the {\em i}-th bin for some values of the parameters $\theta$.  
The minimum of this function follows a chi-square distribution and can be used 
to estimate the goodness of the fit.

The expectation for each reconstructed 
$E_\nu$ and $Q^2$ bin is computed as follows:
\begin{eqnarray}
\nonumber
N_{total}(n_{track},E_\nu^{rec},Q^2_{rec})  =  A \; \Big[
N_{QE}(n_{track},E_\nu^{rec},Q^2_{rec}) \\
+ \; B \times N_{nonQE}(n_{track},E_\nu^{rec},Q^2_{rec}) \Big] 
\; \; \;
\label{Eq.expectedevents}
\end{eqnarray}
where $N_{QE}$ and $N_{nonQE}$ are the separate contributions
of quasi-elastic and non quasi-elastic events.

The free parameter A is the overall normalization. Five parameters $\Phi(E_\nu^{true})$,
not included in the above expression,
are used rescale the neutrino flux in each energy region,
four of which are unconstrained in the fit, while the relative flux for energies
from 1.0 GeV to 1.5 GeV is fixed at 1.0.  The flux is reweighted based on the
true energy of the MC events, and applies to both QE and nonQE events.
The nonQE background is reweighted
using the unconstrained parameter B, which is referred in the rest of this paper
as the ratio nonQE/QE: the relative reweighting of our default MC calculation.
Because of the separation of the two-track QE and non-QE samples,
the nonQE/QE ratio will be constrained by the background and allow 
a fit for the QE axial-vector form factor. 
The parameters A and B are relative to the data/MC normalization 
calculated using all other parameters at their nominal values,
including $M_A$ = 1.1 GeV. Importantly, changing the value 
for $M_A^{QE}$ changes the absolute cross-section for QE, which will in turn
affect the fit value for the free nonQE/QE parameter and the overall normalization.

In this expression, $N_{QE}$ is based on a calculation of the quasi-elastic
cross section with the free parameter $M_A$.  This cross section is 
computed using the true energy and $Q^2$ and convoluted with the 
detailed shape of neutrino energy spectrum, flux(E), from
the beam MC calculations and the hadron production parameterization
used in \cite{K2K:2001}.
\begin{eqnarray}
\nonumber N_{QE}(n_{track},E_\nu^{rec},Q^2_{rec})
 = \!\!\!\!\!\! \sum^{\mathrm{all \; bins}}_{Etrue,Q^2true} \!\!\!\!\!\!
\Big[
\; \; \mathrm{flux}(E_\nu^{true}) \\
\nonumber \times \; d\sigma/dQ^2(E_\nu^{true},Q^2_{true},M_A) 
\; \times \; R(E_\nu^{true},Q^2_{true}) \\
\times \; M(E_\nu^{true},Q^2_{true} \rightarrow n_{track},E_\nu^{rec},Q^2_{rec})\Big].
\end{eqnarray}
Nuclear effects that modify the cross section, especially Pauli blocking
and and other effects of the nucleon momentum distribution, are included using the
factor R, and are discussed in Sec. V.

Because the cross section is calculated using true kinematics, it must 
be modified to account for detector acceptance and resolution, as well 
as nuclear final state interactions, in order to obtain the expectation
in different reconstructed $E_\nu$ and $Q^2$ bins.
This is done with a migration matrix M in the above equation 
where $n_{track}$ refers to the one-track, two-track QE, and two-track
non-QE samples.  This matrix is computed directly from the Monte Carlo simulation.
This result is then applied to the calculated cross section to determine
the number of QE events in each reconstructed $E_\nu$ and $Q^2$ bin.
In contrast, the shape of the non-QE background is taken directly from
the Monte Carlo simulation and already includes these effects.
  
The combination of four flux reweighting
factors $\Phi(E_\nu)$  and the overall normalization are unconstrained.  
The parameter $M_A$ itself affects
the total cross-section as a function of energy.
In this way, we are fitting the shape
of the $Q^2_{rec}$ distribution separately {\em in each energy region}.
This ensures that the axial mass measurement is not significantly biased
by the normalization in any one energy bin.

Finally, the lowest $Q^2_{rec}$ bins, events below 0.2 (GeV/c)$^2$, are not
included in the fit.  The low $Q^2$ region is where there is the largest uncertainty
due to the model for nuclear effects, especially Pauli blocking.  
This eliminates almost half the data, and the total number of events actually included
in the fit is shown in the second column for each data set in Tab.~\ref{Tab.Events}.
Low $Q^2_{rec}$ events are also events at low angle, shown by the cos($\theta$)
term in parenthesis in Eq. \ref{Eq.Q2}, and corresponds to the right-most two bins in the 
cos($\theta$) histograms in Fig.~\ref{Fig.lowQ2} for neutrino energies around 1.0 GeV.

\subsection{Fit Parameters}

We fit a large collection of $E_\nu^{rec}$ and $Q^2_{rec}$ distributions:
two data sets K2K-I and K2K-IIa, each with one-track, two-track QE,
and two-track non-QE subsamples, a total of 242 bins.
The Monte Carlo predictions for these data sets are computed separately
using MC samples that are more than 15 times larger than the data.
The free parameters $\Phi$ for the flux at each energy are common to both
data sets, as is the overall normalization factor,
the non-QE/QE ratio and proton rescattering.  
There are separate 2-track to 1-track
migration parameters for each data set, ten parameters in total.

These last three parameters are constrained by adding a term to the chisquare.
A reweighting or migration is computed and a systematic error for each of these,
assumed to be approximately Gaussian, is estimated from studies of the detector 
and interaction Monte Carlo simulations.  As the fit is performed, a chisquare
is evaluated, assuming this Gaussian shape and error, and is added to the total 
chisquare.  The total degrees of freedom is thus (242 + 3) chisquare terms - 10 
parameters = 235 degrees of freedom.

Proton rescattering
is taken to be uncertain by $\pm$ 20\% from the nominal value used in the NEUT
Monte Carlo simulation.  A reweighting of events is calculated from a second 
full-detector Monte Carlo simulation using 80\% of the proton reinteraction cross sections,
and the systematic error parameter is used to interpolate between these data samples.  
This reweighting has the primary effect
of increasing or decreasing the number of QE events in the two-track sample, by up to 
+4.5\% for the 80\% case. The actual
effect on the fit is to increase or decrease the number of events in the QE enhanced sample
because only that part of the two-track sample has high QE purity.


\begin{figure*}[htbp!]
\begin{center}
\includegraphics[width=5.6cm]{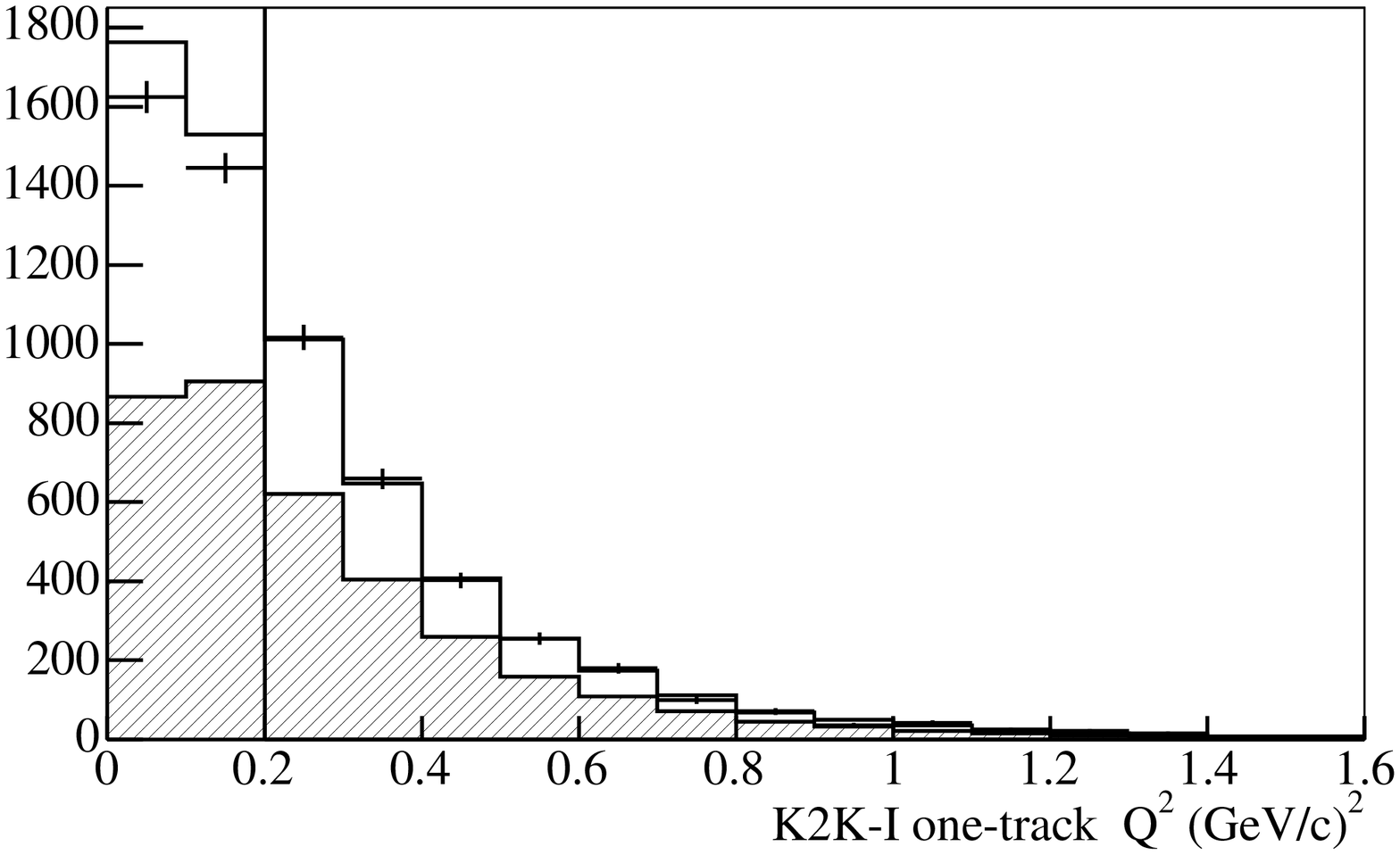}
\includegraphics[width=5.6cm]{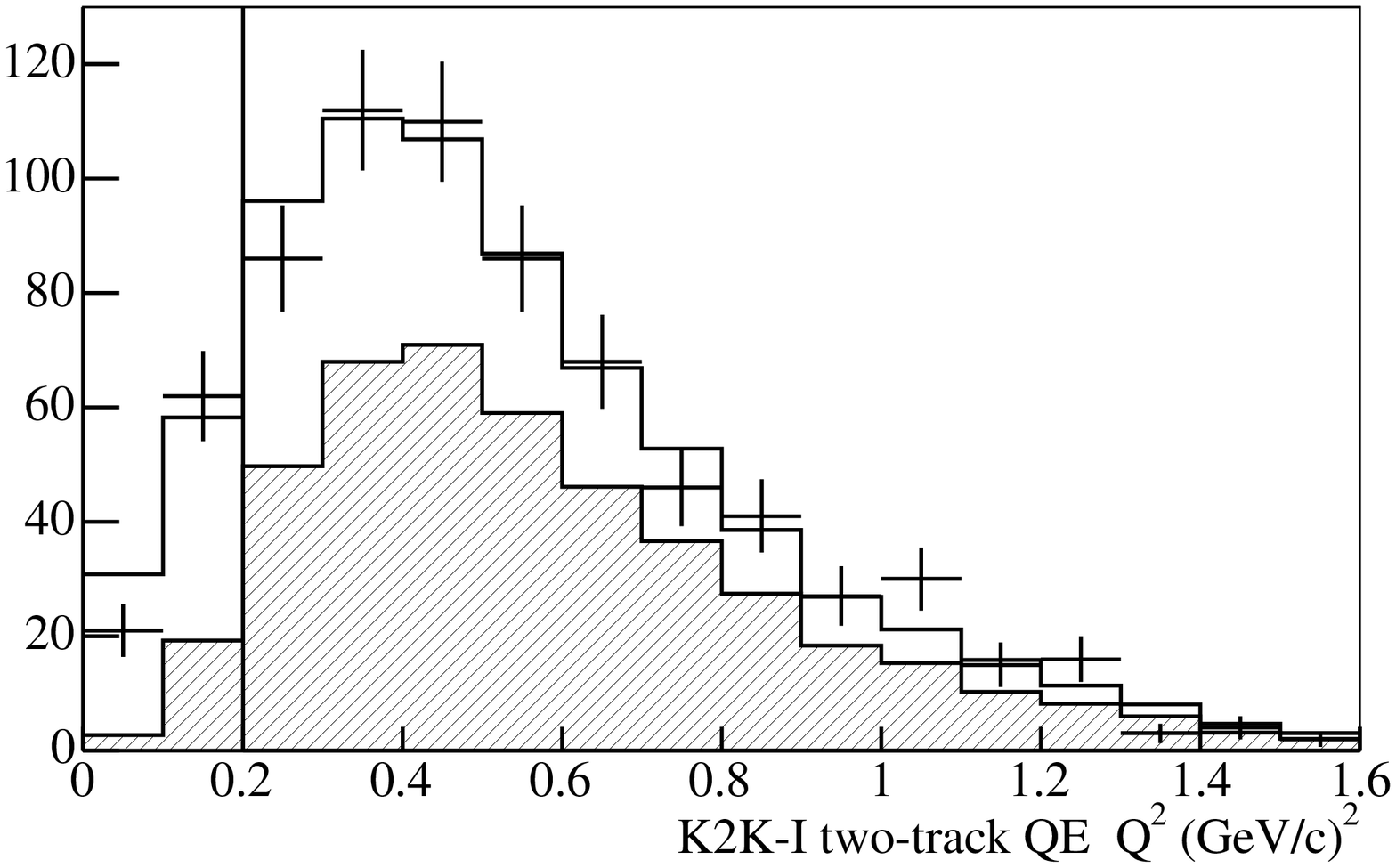}
\includegraphics[width=5.6cm]{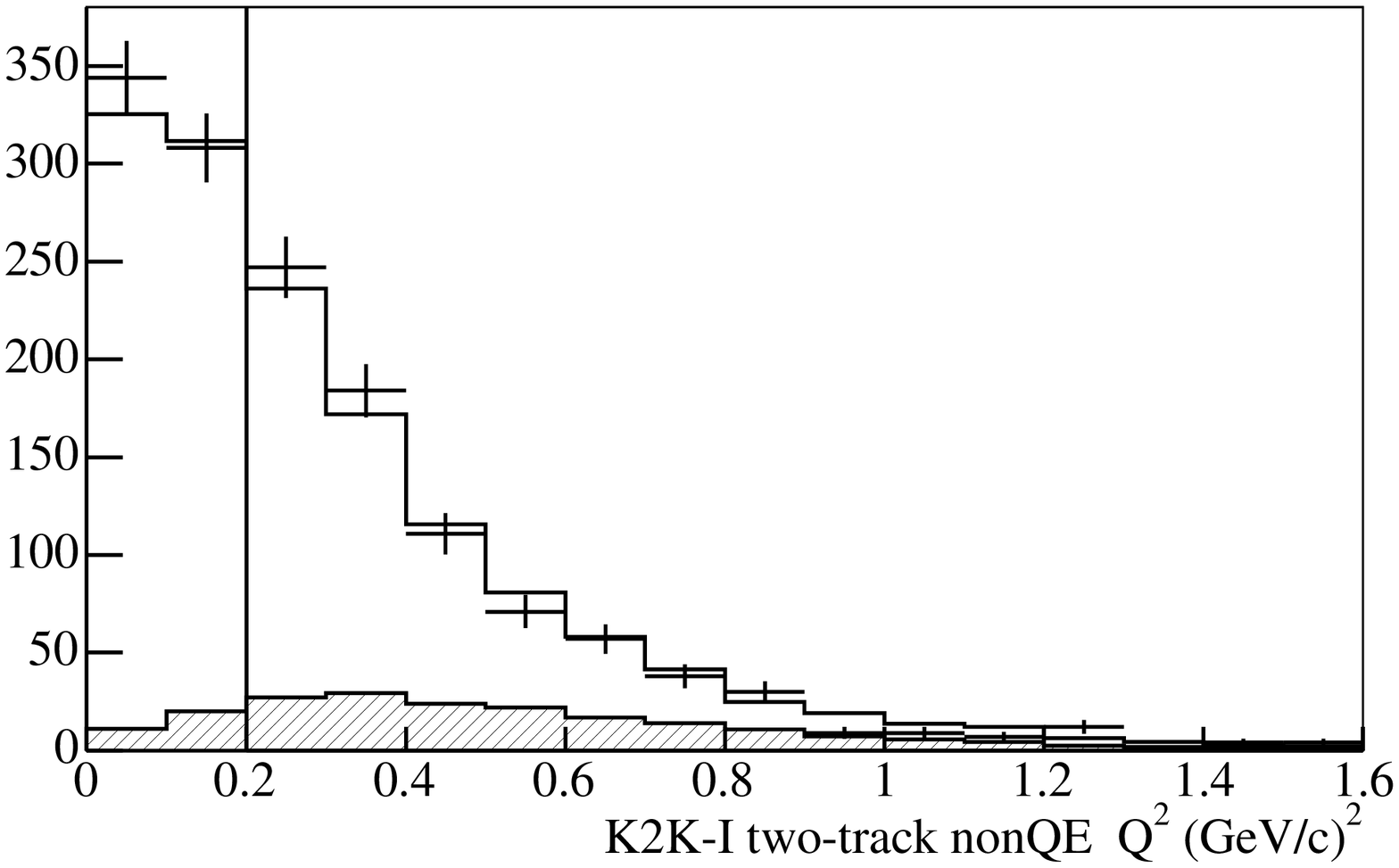}
\includegraphics[width=5.6cm]{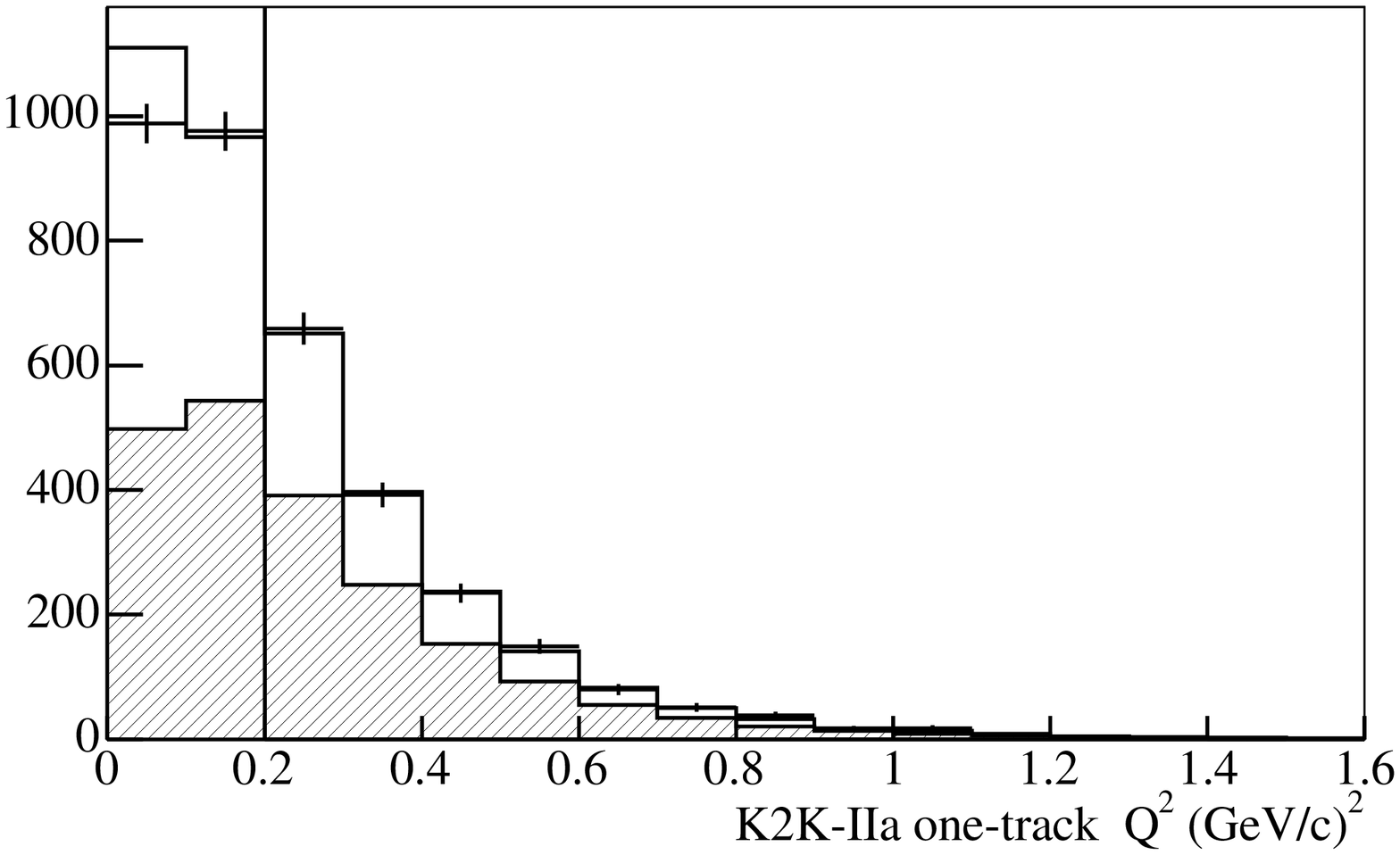}
\includegraphics[width=5.6cm]{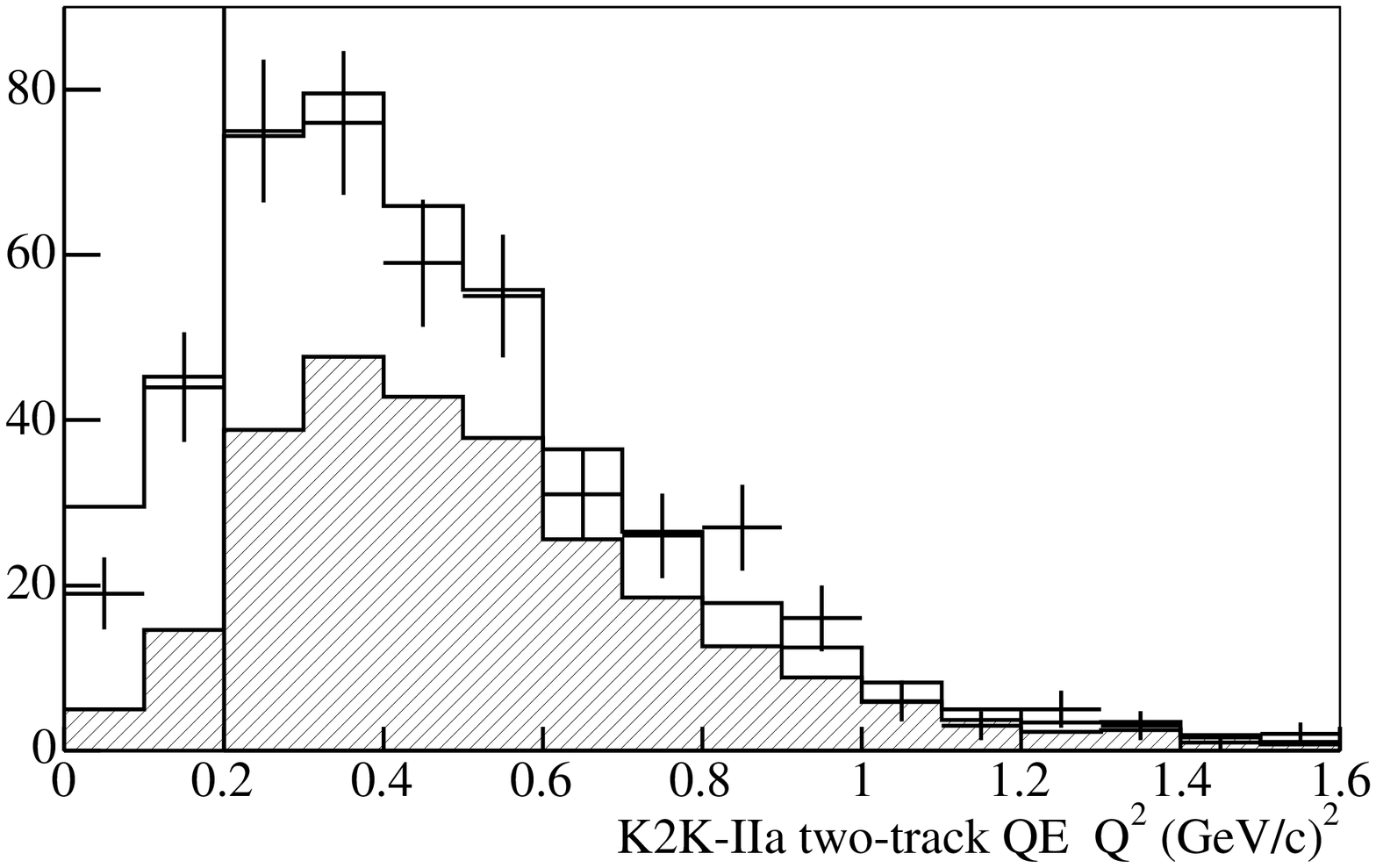}
\includegraphics[width=5.6cm]{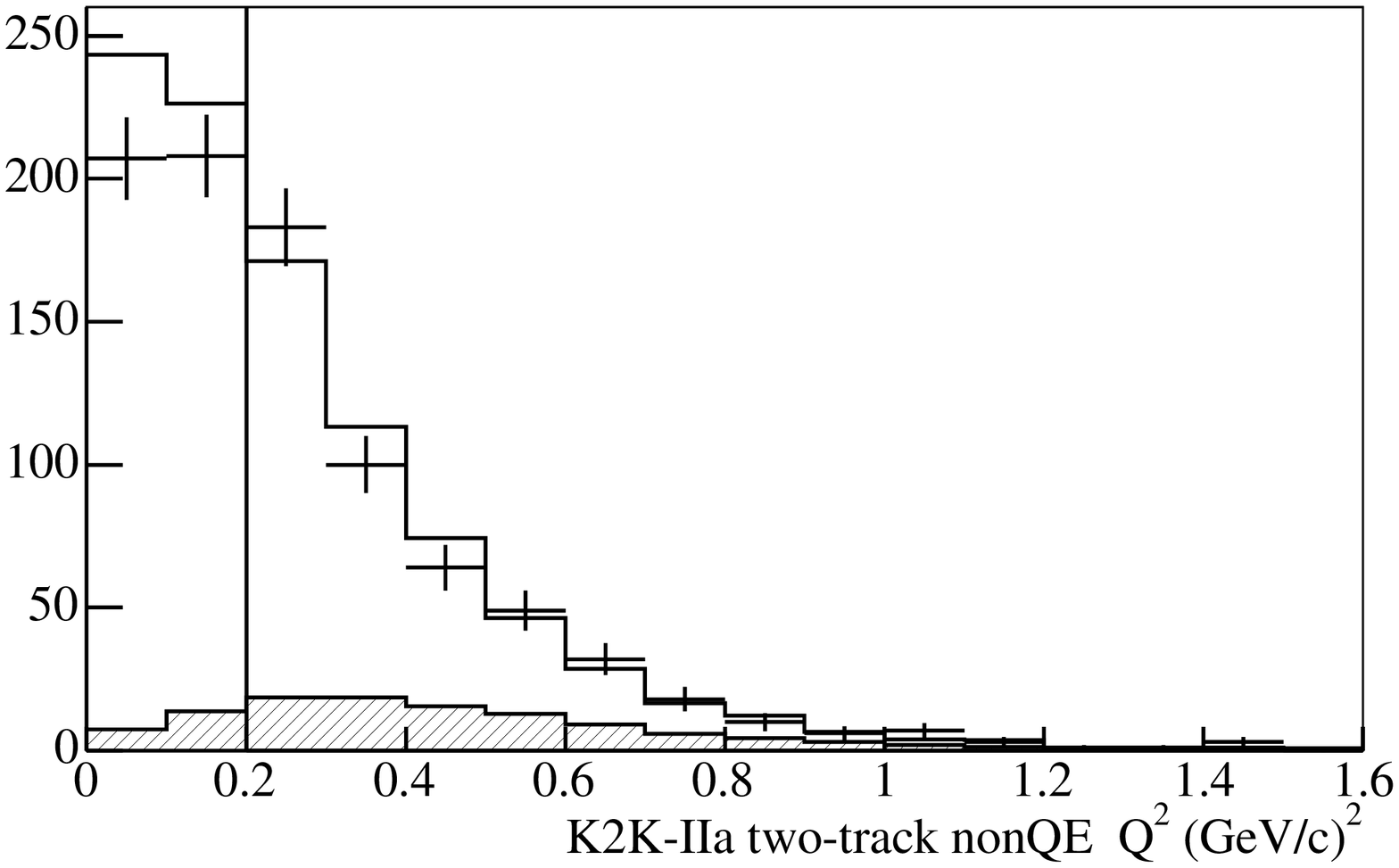}

\caption{The data and the best fit $Q^2_{rec}$ distributions for K2K-1 data (top)
and K2K-IIa data (bottom) for the 1-track, 2-track QE enhanced, and 2-track
non-QE enhanced samples.  The shaded region shows the QE fraction of each
sample, estimated from the MC.  The contribution from each energy region is summed for each plot.
The lowest two data points in each plot are not included in the fit, due to the large
uncertainty in the effects of the nucleus.}
\label{Fig.results}
\end{center}
\end{figure*}

The 2-track to 1-track migration is assigned a $\pm$ 5\% error.
This error is based on the estimated error in the track-finding efficiency
for short, second tracks.
Because the detector response and tracking is tuned separately
for the K2K-I and K2K-IIa data samples, we include separate parameters for each sample
in our fit.
This migration is applied to all events in the MC, not just the QE events
as for proton rescattering.
This parameter is also 
correlated with neutrino interaction effects such as proton and pion rescattering.

%% file: results.tex
\section{Results and Discussion}


\subsection{Fit Results}

The result of the combined fit is $M_A$ = 1.20 $\pm$ 0.12 GeV.
The chisquare value for this fit is 261 for 235 degrees of freedom.
The $Q^2$ distributions for the data and the MC simulation with the 
best fit $M_A$
are shown in Fig. \ref{Fig.results}, with all five energy regions
combined.  
The best fit values for the free parameters
in the fit are summarized in Tab. \ref{Tab.fitvalues}.

In the fit, there is a strong correlation between the
normalization, nonQE/QE, and the two-track to one-track 
migration, but different ways of constraining the parameters
do not affect the fit value of $M_A$ very much.
The migration is expressed such that 0.90 means 10\% of the two-track
events in each $E_{rec}$ and $Q^2_{rec}$ bin should be moved 
to the corresponding one-track bin; this parameter is being pulled
beyond its 5\% gaussian constraint for both samples.  
For example, fixing the 2-track to 1-track migration 
at 0.95, the one-sigma edge of the 5\% error, 
yields normalization = 0.975, nonQE/QE = 1.26,
and proton rescattering 1.20.
However, $M_A$ is unchanged while the chisquare rises to 261
for 234 degrees of freedom. 

Finally, one further comment about the nonQE/QE parameter.  The higher
fit value for $M_A^{QE}$=1.20 GeV causes an increase of about 10\% in the absolute
QE cross section, relative to the default value of $M_A^{QE}$=1.10 GeV.  
If a good fit requires maintaining a similar relative non-QE cross-section,  
then a corresponding increase in nonQE/QE
parameter and decrease in the absolute normalization is required.
This appears as a portion of the nonQE/QE = 1.30 fit value and normalization.

\begin{table}[htpb!]
\begin{tabular}{c|cc}
parameter & fit value & error 
\\ \hline $M_A$ (GeV)     & 1.20 & 0.09
\\ $\Phi$(0.5 to 1.0 GeV) & 1.02 & 0.25
\\ $\Phi$(1.0 to 1.5 GeV) & 1.00 & fixed
\\ $\Phi$(1.5 to 2.0 GeV) & 0.80 & 0.09
\\ $\Phi$(2.0 to 2.5 GeV) & 0.93 & 0.08
\\ $\Phi$($>$ 2.5 GeV)    & 1.09 & 0.11
\\ Normalization & 0.96 & 0.09
\\ nonQE/QE            & 1.30 & 0.17
\\ K2K-I 2tk $\rightarrow$ 1tk & 0.94 & 0.02
\\ K2K-IIa 2tk $\rightarrow$ 1tk & 0.94 & 0.03
\\ Proton Rescattering & 1.14 & 0.18
\end{tabular}

\caption{Best fit values for the parameters in the fit.
The errors given are from the fit only.  
The error for $M_A$ rises to $\pm$0.12 GeV when
the other systematic effects are included}
\label{Tab.fitvalues}
\end{table}

\subsection{Consistency checks}

\begin{figure}[htbp!]
\begin{center}
\includegraphics[width=9cm]{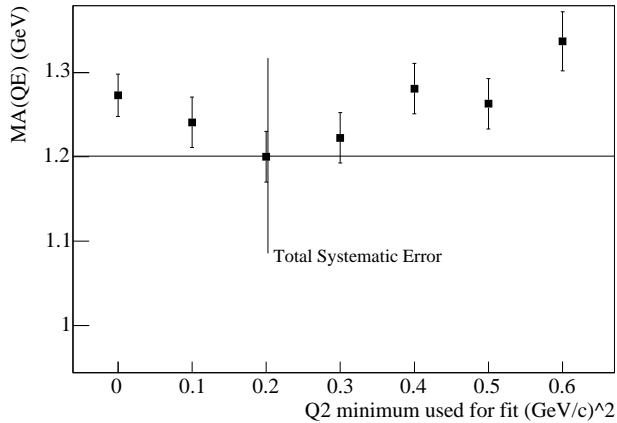}

\caption{Fit values obtained for different values of the low $Q^2$
cut.  Only statistical errors are shown.
The horizontal line is the combined best fit.  The vertical line is 
the systematic errors.  
}
\label{Fig.q2cut}
\end{center}
\end{figure}

A test for consistency is to vary the low $Q^2$ cut and compare the results.  
In Fig. \ref{Fig.q2cut}, the best fit $M_A$ is shown with different
minimum $Q^2_{rec}$.  The error bars include an estimate of statistical 
errors only; however, the data for each point in this figure are correlated
with the other points.
The extra error bar shows the total error.
When no cut is applied (and no coherent pion), 
the fit value is
$M_A$ = 1.27 $\pm$ 0.12 GeV, 
where the statistical error is less, but a large additional systematic
error of $\pm$ 0.07 GeV is assigned (but not shown in Fig. \ref{Fig.q2cut}) 
due to uncertainty in the amount of Pauli blocking.  When the systematic
errors are considered, these results are consistent with the fit value.

\begin{figure}[htbp!]
\begin{center}
\includegraphics[width=9cm]{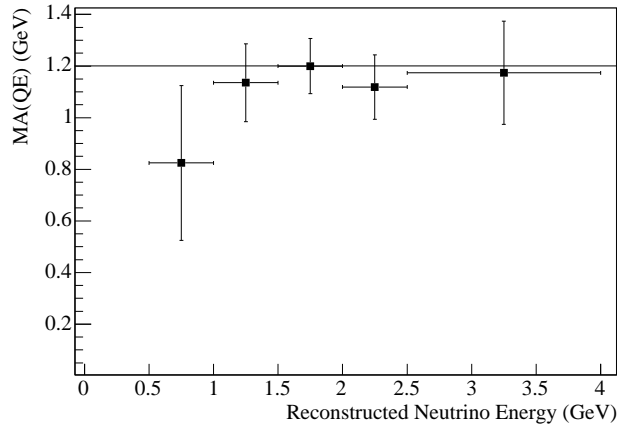}

\caption{Fit values obtained separately from the shape of the $Q^2$
distribution for each neutrino energy.  The horizontal line indicates
the combined best fit value.
}
\label{Fig.energy}
\end{center}
\end{figure}

A second check is to consider the fit values for the $Q^2$
distribution {\em at each energy}, shown in Fig. \ref{Fig.energy}.  
This uses the best fit values for the flux for all energies except the 
one being tested while the chisquare, and therefore the shape fit, 
is computed only for the energy bins in question.  This is necessary
because of the significant migration from true energy 
(where the flux parameter is applied)
to reconstructed energy used in the fit, especially for the non-QE background.
There are different systematic effects, and this 
result should not be considered a measurement, but rather a consistency test.
However results for each energy are also consistent with the combined result.

Finally, we fit for the value of $M_A$ for the K2K-I and K2K-IIa data sets separately,
and obtain the values 1.12 $\pm$ 0.12 ($\chi^2$ 150/127 dof) and 1.25 $\pm$ 0.18 GeV
($\chi^2$ = 109/101 dof) respectively.  The primary 
difference between the two data sets are the presence or absence of the lead glass detector
and the accompanying uncertainty in the muon momentum scale.  Also, the muon thresholds are 
somewhat different, and the acceptance model for short, second tracks in SciFi is calibrated
 separately for the two data sets, which could show up in the 2-track to 1-track migration.

\subsection{Systematic uncertainties}

Here these systematic errors are discussed in detail.
The largest contributions to the systematic error, summarized in 
Tab.~\ref{Tab.errors},
are the uncertainty
in the muon momentum scale, and the normalization and uncertainty in the flux for each
energy region.  Other, smaller contributions include the shape of the
non-QE background, the non-QE/QE ratio, and the two-track to one-track migration. 
A final, interesting source
of uncertainty comes from nuclear effects, though it contributes only
a small amount to this analysis.  The statistical error is estimated
by setting all the other parameters in the fit to their best fit values
and determining the resulting error in $M_A$, though there is a further
statistical effect in the normalization parameter.

\begin{table}[ht]
\begin{tabular}{lc|r}
Sources of uncertainty & & Error in MA
\\\hline Muon momentum scale & & 0.07
\\Relative Flux and normalization & & 0.06
\\ $M_A$ 1-$\pi$ & & 0.03
\\nonQE/QE & & 0.03
\\Proton rescattering & & 0.03
\\Statistics & & 0.03
\\\hline Total & & 0.12 
\end{tabular}

\caption{The calculation of the total error.  
Errors smaller than 0.03 are not listed. 
The total value takes into account the correlations among those 
errors that are parameters in the $M_A$ fit; 
the others are added to that total in quadrature.
}
\label{Tab.errors}
\end{table}

\subsubsection{Muon momentum scale}

The muon momentum appears directly, and indirectly via $E_\nu$, in 
the calculation of the value of $Q^2_{rec}$ for each event.  The uncertain
absolute scale for this momentum, as modeled in the detector Monte Carlo simulation, 
will cause the MC prediction for the shape of the $Q^2_{rec}$ distribution to 
be more or less compressed and affect the $M_A$ measurement.  
As an example, a $\pm$ 1\% error in the 
momentum scale gives a $\mp$ 0.05 GeV error in the fit value for $M_A$.
Approximately $\mp$ 0.01 GeV of this error can be attributed to shifting a small 
number of events up or down one $E_\nu^{rec}$ bin.
The other $\mp$ 0.04 GeV is from the calculation of the reconstructed
$Q^2$ itself.  The central value for the muon momentum scale
is determined from the neutrino energy spectrum analysis, described in Sec. III 
and reference\cite{K2K:2005}, while
the error from that analysis is propagated to the $M_A$ analysis as described below.

Because the muon momentum is measured using its range in the detector,
the uncertainty for the overall momentum may come from any of the pieces of the detector:
SciFi, LG, SciBar, or the MRD.  In this analysis, we model this uncertainty
by assigning it to just two of these.  The first is the uncertainty in the
density of the lead-glass detector and therefore the energy loss
experienced by the muon passing through it.  The second piece is a
scaling factor for part of the muon momentum calculated from the range
in the MRD detector.   For both pieces, we determine the central
value of the momentum shift and the error from the neutrino data.

The density of the lead glass, which is incorporated into the geometry
description in our Monte Carlo simulation, is determined from a beam test
and is uncertain by 5\%.
We have modeled the effect of this uncertainty and made a reweighting
table that modifies the MC p$_\mu$ and $\theta_\mu$ distribution.
This uncertainty could give rise to a 2\% error in the total momentum for
a typical K2K-I event.  In the neutrino energy spectrum measurement, 
this is a parameter in the fit and good
agreement with the data is found with a value that is 0.98 $\pm$
0.013 times the density obtained from the beam test; the neutrino data
provide the stronger constraint.  This central value is used in the
$M_A$ analysis.

Likewise, we measure a shift in the muon momentum scale for the
Muon Range Detector of 0.976 $\pm$ 0.007 using the energy spectrum measurement
procedure.  When the K2K-I and K2K-IIa data are fit separately, we 
obtain a consistent result for this parameter, despite the presence of
the lead glass detector in the former.
This is assigned as an error for the MRD portion of the muon range, but it
actually arises from a
combination of factors including the material assay for the MRD and
SciFi (about 1\%), the simulation of muon energy loss in GEANT
\cite{GEANT3} (about 1\%) and the uncertainty in the intrinsic muon momentum from the
neutrino interaction MC (about 0.5\%).  Again, we find the neutrino
data produce a good central value and a tighter constraint than
taking the individual errors in quadrature.
Though these errors actually come from all portions of the muon track,
we find no significant difference in the analysis if this
factor is obtained from and applied to the whole track momentum, instead of the MRD
portion only.

Because the $M_A$ fit and the energy spectrum analysis use the same neutrino data,
it is possible that the uncertain value for $M_A$ itself is affecting
the fit values for the MRD muon momentum scale when that value is obtained
from the spectrum measurement.   Our default Monte Carlo assumes $M_A$ = 1.1
GeV.  An uncertainty in this value of $\pm$ 0.20 GeV corresponds to an
error of $\pm$ 0.01 in the fit value of the momentum scale.  
This is taken as an additional uncertainty when this
parameter is used to determine $M_A$.  Also, there is a
correlation between the lead-glass density error and the MRD momentum
error.  When all of these effects are combined, the resulting error in
$M_A$ is $\pm$0.07 GeV.

\subsubsection{Flux for each energy region and normalization}

A significant uncertainty in $M_A$ arises because the relative neutrino flux for 
each energy region and the overall normalization parameters 
are unconstrained parameters in the fit.
In this way we are fitting the shape of the $Q^2$ distribution in each energy
region separately, regardless of the errors in the incident neutrino flux.
The contribution to the total error is estimated by fixing the other 
free parameters in the fit and reading the resulting error in $M_A$,
which is from these parameters and statistics only.

The overall normalization contributes more to the error than the 
uncertain relative normalization.  This is estimated by further constraining
the relative flux so that only the normalization and $M_A$ are free.  
The overall normalization
is correlated with $M_A$ because $M_A$ affects both the 
relative size of and the shape of the QE cross-section.  Different
combinations of $M_A$ and normalization will give a reasonable chisquare
when compared with the data,
and the error due to this parameter, more than the others, 
would be reduced with increased data statistics, even with no further
constraints.  This last result is confirmed using MC samples of various
sizes as if they were data, to study the effect of statistics of the data sample.

We do have a constraint on the relative flux for each energy region from the 
neutrino oscillation measurement \cite{K2K:2005}.  This measurement is done
using data from all the near detectors, not just SciFi.  This information
is not completely independent of this analysis because it shares some
of the same data set, but a different analysis technique, and several other
data sets from the other near detectors.
We get a consistent result $M_A$ = 1.13 $\pm$ 0.12 GeV when this 
constraint is used.

\subsubsection{nonQE/QE parameter and two-track to one-track migration}

The nonQE/QE scaling ratio is also a free parameter in the axial-mass fit.  
There is no constraint on this parameter for this analysis, though
other estimates find that it is uncertain by 5 to 10\% \cite{K2K:2005}.
The two-track to one-track migration parameter is highly correlated with
nonQE/QE, and when these two are combined, they contribute a total error
of 0.03 GeV to the fit value of $M_A$.  As before, this is obtained by fixing
all the other parameters such that the resulting error in $M_A$ is the combination
of these and the statistical error only.

\subsubsection{Non quasi-elastic background shape}

Single pion events from the production and decay of the $\Delta$ 
and other resonances in the nucleus are the largest background 
to the QE samples in this analysis.
These events are described by a calculation that 
includes a similar axial mass parameter which affects
the shape of the $Q^2$ distribution.  If the value used to model
the single pion background is different, that will affect the 
fit value obtained for the quasi-elastic events.  Our calculation
takes $M_A^{1\pi}$ = 1.1 $\pm$ 0.1 GeV.  This contributes an 
uncertainty of $\pm$ 0.03 to result for $M_A^{QE}$, and is estimated
by generating a second complete MC sample with $M_A^{1\pi}$ = 1.2 GeV.

Other contributions to the nonQE background are deep inelastic scattering
and coherent pion production.  For the former, we have evaluated the
uncertainty by removing the Bodek-Yang correction, and find no effect.
We also consider the case where charged-current coherent pion events 
are produced which has only +0.01 GeV effect for the $Q^2 > 0.2$ cut
used in the standard analysis, but increases the fit value by 0.10 GeV 
when we fit the entire $Q^2$ range.

\subsubsection{Nuclear effects}

\begin{figure*}[htbp!]
\begin{center}
\includegraphics[width=8cm]{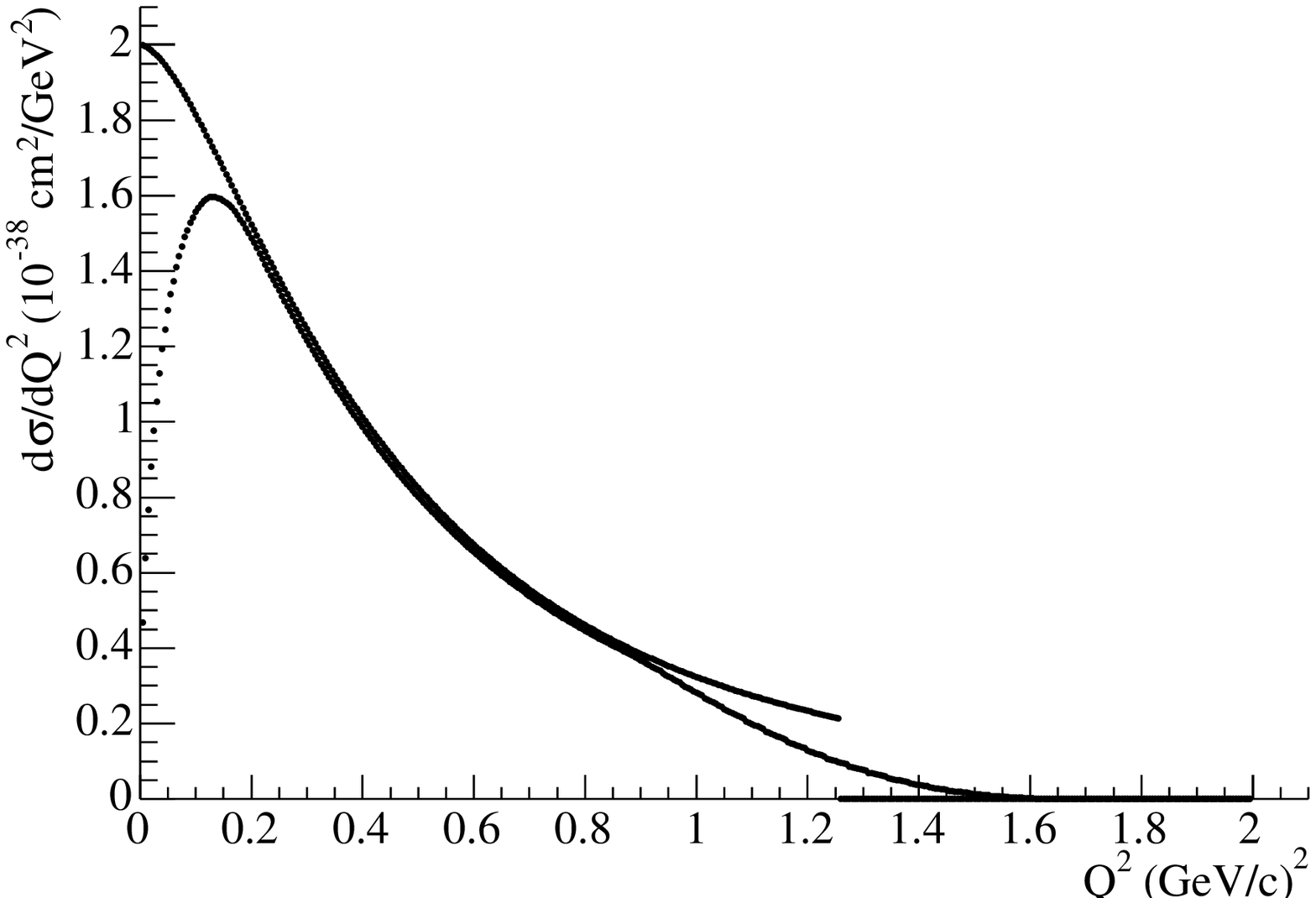}
\includegraphics[width=8cm]{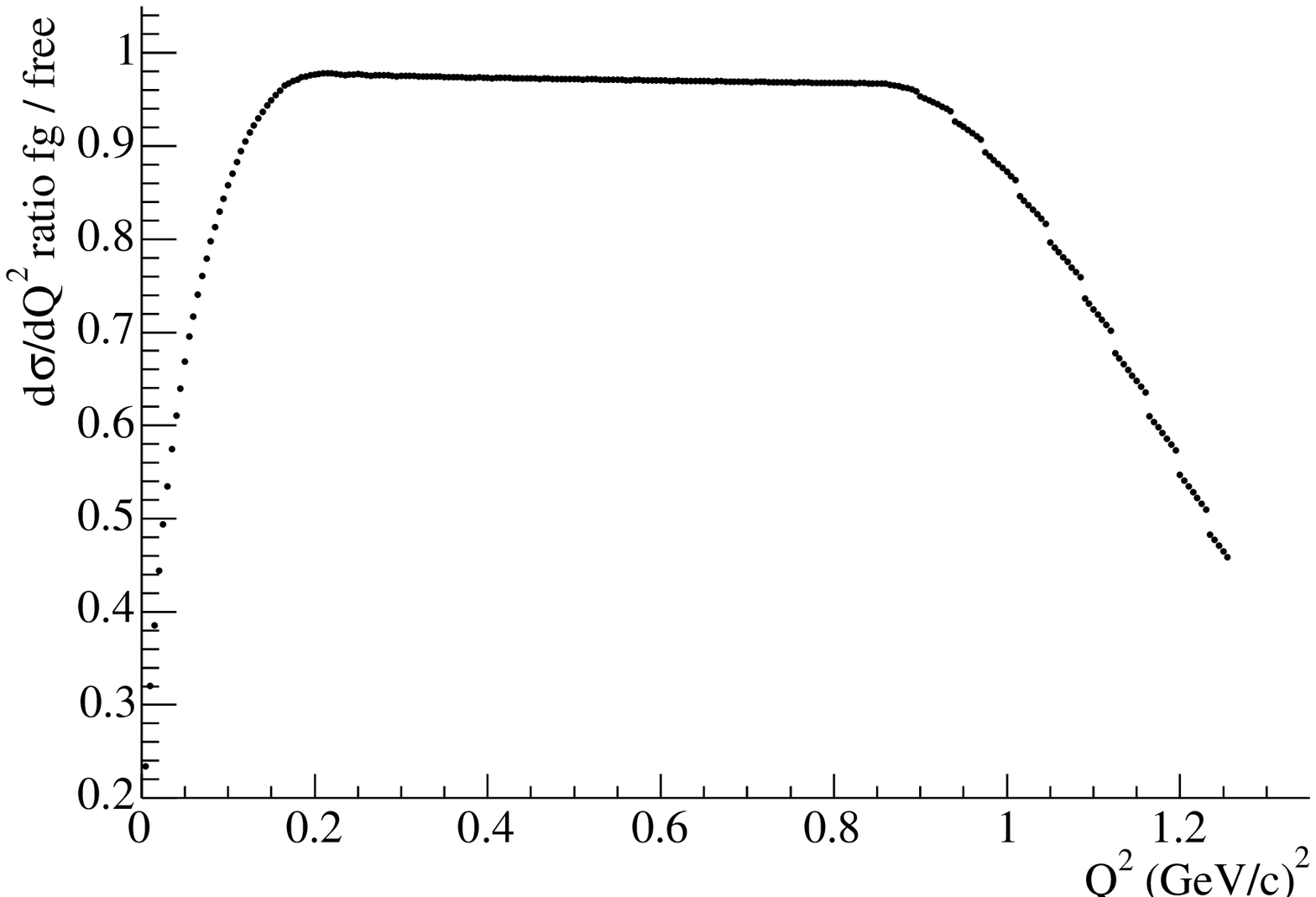}

\caption{Effect of the nucleon momentum on the shape of the $Q^2$
distribution.  The comparison is between the free nucleon and a uniform Fermi
gas model.  The effect of Pauli blocking is seen at low $Q^2$, the
tail of the momentum distribution at high $Q^2$, an overall suppression,
and a slight change in the slope in the middle region.
The calculated quasi-elastic cross sections for 1.0 GeV neutrinos on oxygen
are on the left, and the ratio (Fermi gas)/ (free neutron).}
\label{Fig.fermigas}
\end{center}
\end{figure*}

Other interesting sources of uncertainty are the effects of the 
nucleus on the cross section and the $Q^2$ distribution, primarily
from the nucleon momentum distribution.   The effects are small
relative to the other uncertainties described above because the 
minimum $Q^2_{rec}$ cut eliminates the data where these errors
are most significant.
These effects will be of interest for future precision experiments
and as models of neutrino-nucleus interactions become more sophisticated.
We present a description of these effects for the uniform Fermi gas model,
in this case from the calculation in \cite{Nakamura:2001,Nakamura:2002}.
The three effects are described below and summarized in 
Fig. \ref{Fig.fermigas} for a 1.0 GeV neutrino.  
It is the ratio in this figure that is the 
basis for R($E_\nu$,$Q^2$) in Eq. \ref{Eq.expectedevents}.

The main
uncertainty is the amount of Pauli blocking that should be applied
both to the quasi-elastic and also the single pion background.
Within the context of the Fermi gas model, 
this can be estimated by assuming a different $k_f$:  215 and 235
in addition to the default value of 225 GeV/c.  
The effects of this uncertainty on $M_A$ do not appear
with the $Q^2 > 0.2$ requirement used in this analysis, but are
as much as 5\% for fits that use the lowest $Q^2$ events.  

At the upper end of the $Q^2$ distribution, the quasi-elastic cross
section has a kinematic cut off whose location depends on the incident
neutrino energy.  The momentum distribution 
in a nucleus smears this step, giving a tail to the distribution.
These high $Q^2$ interactions produce muons that do not reach the MRD
because they are at high angle or their momentum is too low,
so this has no effect on the present analysis.

The momentum distribution will modify the shape of the
$Q^2$ distribution through the middle region between the two effects
described in the preceding paragraphs.  The slope of the middle
region in the second plot in Fig. \ref{Fig.fermigas} is approximately
0.017 (GeV/c)$^{-2}$.  There is also an overall suppression of the cross
section of 2\%.  The uncertainty represented by the change in slope can be
propagated to the $M_A$ analysis by modifying R(E,$Q^2$) in the fit.  
The resulting uncertainty in $M_A$ is $\pm 0.01$ GeV,
negligible compared to the other uncertainties in this analysis.

A final uncertainty from the nuclear model is the nucleon interaction
energy.  For our Fermi gas model, this takes the form of an effective
binding energy -27 $\pm$ 3 MeV,
and is the energy given up to the recoil proton from the nucleus.
This affects the outgoing muon momentum and would contribute $\pm$ 0.02 GeV error
to $M_A$, but this is naturally included by the free muon momentum scale parameter
in this analysis.

These uncertainties are also used to estimate the effect of the 21.8\%
aluminum that makes up the fiducial mass. 
The neutrino-aluminum interactions are
taken to have the same cross section per nucleon and the same 
kinematics as for oxygen.  A higher $k_f$ appropriate for aluminum
only has an effect in the
Pauli blocked region. The increased effective binding energy is
equivalent to a shift in p$_\mu$ of about 3 MeV for this fraction
of the interactions, and thus is negligible for the whole sample.

\begin{table*}[ht]
\begin{tabular}{lc|cc|ccc}
\\ Experiment & Pub. Date & Target & Method & $M_A$ & Error & comment
\\\hline ANL \cite{ANL:1982} & 1982 & D & 12' Bubble Chamber & 1.00 & $\pm$ 0.05 &
\\ FNAL \cite{Fermilab:1983} & 1983 & D & 15' Bubble Chamber & 1.05 & +0.12 - 0.16 &
\\ BNL \cite{BNL:1990} & 1990 & D & 7' Bubble Chamber & 1.07 & +0.040 -0.045 &
\\ CERN \cite{GGM:1977} & 1977 & CF$_3$Br & GGM Bubble Chamber & 0.94 & $\pm$ 0.17 &
\\ CERN \cite{GGM:1979} & 1979 & CF$_3$Br, C$_3$H$_8$ & GGM Bubble Chamber & 0.94 & $\pm$ 0.05 &
\\ SKAT \cite{SKAT:1990} & 1990 & CF$_3$Br & Bubble Chamber & 1.05 & $\pm$ 0.14 & ($\nu$) 
\\ SKAT \cite{SKAT:1990} & 1990 & CF$_3$Br & Bubble Chamber & 0.79 & $\pm$ 0.20 & ($\bar{\nu}$)
\\ BNL \cite{BNL:1969} & 1969 & Fe & Segmented Tracker & 1.05 & $\pm$ 0.20 & 
\\ BNL \cite{BNL:1987} & 1987 & HC, Al & Segmented Tracker & 1.06 & $\pm$ 0.05 & elastic scattering
\\ BNL \cite{BNL:1988} & 1988 & HC, Al & Segmented Tracker & 1.09 & $\pm$ 0.04 & ($\bar{\nu}$)
\\ K2K SciFi & this expt. & H$_2$O, Al & Segmented Tracker & 1.27 & $\pm$ 0.12 & dipole form factors
\end{tabular}
\caption{Results from other experiments, grouped first by target nucleus, then by publication date.  
Where separate values are given
for $M_A$ extracted from the shape of d$\sigma$/d$Q^2$ only, that is the value
included in this table.  All the data are for the neutrino quasi-elastic reaction 
($\nu \; n \rightarrow \mu^- \; p$) except for two which also took data with 
anti-neutrino ($\bar{\nu}\; p \rightarrow \mu^+ \; n$), one of which studied 
neutral current (elastic) scattering, noted in the table.  
For better comparison with other experiments, the K2K SciFi result is the one analyzed 
with dipole vector form factors.
}
\label{Tab.otherexperiments}
\end{table*}

\subsection{Effect of the new vector form factors}

The basic method used to measure the axial vector mass here is the same
as for previous measurements, listed in Tab. \ref{Tab.otherexperiments}, 
but since that time there have been 
improved measurements for the shape of the vector form factors
from electron scattering experiments.  Changing the shape of the 
contribution of the vector form factors affects the fit shape of the 
axial-vector form factor. We continue to assume $F_A$ has a 
dipole form; future high-precision neutrino experiments may be sensitive
to subtle differences from a $F_A$ dipole shape.
  
Our results assume  the updated parameterizations 
of Bosted \cite{Bosted:1995}.  We have evaluated one other recent
parametrization \cite{BBA:2002} and found the $M_A$ result differs by only 0.01 GeV.
This is true even considering the discrepancy between
the polarization transfer measurement and the Rosenbluth separation measurement,
described in \cite{BBA:2002} with further references.  
For our analysis, we use the parameterization of $G^N_E$ given by 
\cite{Galster:1971}.

In order to allow comparison with the previous results, we have repeated
the analysis with the same modified dipole approximations used by 
\cite{BNL:1990,Fermilab:1983,ANL:1982} who follow
Olsson {\it et al.} \cite{Olsson:1978}.  
We find that these old 
parameterizations produce a value that is 1.23, roughly 0.03 higher.
When only a pure dipole is used, 
the fit value is 1.27 GeV.

\subsection{Comparison with other experiments}

This is the first measurement of the axial vector mass using neutrino
interactions with oxygen targets,
but there have been many previous measurements with a variety of other
target nuclei.  The experiments in Tab. \ref{Tab.otherexperiments} 
have hundreds or thousands
of events from neutrino or anti-neutrinos with energies of a few GeV.
The systematic errors they report are dominated by uncertainties 
in the neutrino flux,
calculation of nuclear effects, and subtraction of non-quasielastic 
backgrounds.

One problem with comparing the results in Tab. \ref{Tab.otherexperiments}
is that the older analyses used not only different assumptions about the
vector form factors, but also different backgrounds 
and other physical constants such as $F_A$($q^2$=0).
The results 
given here are the published values; however, the authors of \cite{BBA:2002}
have made some effort to reproduce and then update all of the analysis 
assumptions for a selection of these experiments.

%% file: conclusion.tex
\section{Conclusion}

We have made the first measurement of axial vector form factor
using neutrino interactions on an oxygen target.  We find
an axial vector mass $M_A$ = 1.20
$\pm$ 0.12 GeV gives the best agreement with the data,
if we assume a dipole form for $F_A$. 
This analysis includes the updated (non-dipole) vector form factors obtained 
from electron scattering experiments.  In order to better compare 
with previous experiments, an alternate result using only pure 
dipole vector form factors is $M_A$ = 1.27 $\pm$ 0.12 GeV.  
We note that this analysis is very sensitive to the absolute muon momentum
scale.
We have also studied the details of the nucleon momentum distribution 
for oxygen on this analysis and find only a small effect on the shape 
of the $Q^2$ distribution for $Q^2 > 0.2$ (GeV/c)$^2$.

%% file: acknowlegements.tex
We thank the KEK and ICRR directorates for their strong support
and encouragement.
K2K is made possible by the inventiveness and the
diligent efforts of the KEK-PS machine group and beam channel group.
We gratefully acknowledge the cooperation of the Kamioka Mining and
Smelting Company.  This work has been supported by the Ministry of
Education, Culture, Sports, Science and Technology of the Government of
Japan, 
the Japan Society for Promotion of Science,
the U.S. Department of Energy,
the Korea Research Foundation,
the Korea Science and Engineering Foundation,
NSERC Canada and Canada Foundation for Innovation,
the Istituto Nazionale di Fisica Nucleare (Italy),
the Spanish Ministry of Science and Technology,
and Polish KBN grants: 1P03B08227 and 1P03B03826.